\documentclass[showpacs,aps]{revtex4-2}
\emergencystretch=\maxdimen
\usepackage{dcolumn}
\usepackage{amsmath,amssymb,mathrsfs,tensor}
\usepackage[latin1]{inputenc}
\usepackage{graphicx, psfrag}
\usepackage[colorlinks=true, citecolor=blue, urlcolor = blue, linkcolor= red, bookmarks=true]{hyperref}
\usepackage{float}
\usepackage{amsfonts}
\usepackage{dcolumn}
\usepackage{hyperref}
\usepackage{subfigure}
\usepackage{pgfplots}
\usepackage{epstopdf}
\usepackage{booktabs}
\usepackage{gensymb}
\usepackage{multirow}

\makeatletter
\def\btt#1{\texttt{\@backslashchar#1}}
\DeclareRobustCommand\bblash{\btt{\@backslashchar}} \makeatother

\begin{document}
	
\title[]{Photon ring structure of rotating regular black holes and no-horizon spacetimes}
\author{Rahul Kumar$^{a}$}\email{rahul.phy3@gmail.com}
\author{Sushant~G.~Ghosh$^{a,\;b}$} \email{sghosh2@jmi.ac.in}
	
\affiliation{$^{a}$ Centre for Theoretical Physics, Jamia Millia Islamia, New Delhi 110025, India}
\affiliation{$^{b}$ Astrophysics and Cosmology Research Unit, School of Mathematics, Statistics and Computer Science, University of KwaZulu-Natal, Private Bag 54001, Durban 4000, South Africa}
	
\date{\today}

\begin{abstract}
The Kerr black holes possess a photon region with prograde and retrograde orbits radii, respectively,  $M\leq r_p^-\leq 3M$ and  $3M\leq r_p^+\leq 4M$, and thereby always cast a closed photon ring or a shadow silhouette for $a\leq M$. For $a>M$, it is a no-horizon spacetime (naked singularity) wherein prograde orbits spiral into the central singularity, and retrograde orbits produce an arc-like shadow with a dark spot at the center.  We compare Kerr black holes' photon ring structure with those produced by three rotating regular spacetimes, viz. Bardeen, Hayward, and nonsingular. These are non-Kerr black hole metrics with an additional deviation parameter of $g$ related to the nonlinear electrodynamics charge. It turns out that for a given $ a $, there exists a critical value of $ g $, $g_E$ such that $\Delta=0$  has no zeros for $ g >  g_E$, one double zero at $ r = r_E  $ for $ g =  g_E $, respectively, corresponding to a no-horizon regular spacetime  and extremal black hole with degenerate horizon. We demonstrate that, unlike the Kerr naked singularity, no-horizon regular spacetimes can possess closed photon ring when  $g_E< g \leq g_c$, e.g., for $a=0.10M$, Bardeen  ($g_E=0.763332M<g\leq g_c= 0.816792M$), Hayward ($g_E=1.05297M < g\leq g_c = 1.164846M$) and  nonsingular ($g_E=1.2020M < g \leq g_c= 1.222461M$) no-horizon spacetimes have closed photon ring. These results confirm that the mere existence of a closed photon ring does not prove that the compact object is necessarily a black hole. The ring circularity deviation observable $\Delta C$ for the three no-horizon rotating spacetimes satisfy  $\Delta C\leq 0.10$ as per the  M87* black hole shadow observations. We have also appended the case of Kerr-Newman no-horizon spacetimes (naked singularities) with similar features.
\end{abstract}

\maketitle

\section{Introduction}
Circular orbits of photons, the extreme cases of gravitational light bending, play a crucial role in the appearance of black holes and offer a promising tool to probe the strong gravity features. A black hole embedded in a bright background or surrounding accretion flows cast a shadow that is fringed by a characteristic sharp and bright emission ring, known as photon ring, whose structure solely depends upon the spacetime geometry \cite{Johannsen:2015qca}. Synge \cite{Synge:1966}, and Luminet \cite{Luminet:1979nyg} led the study of a non-rotating black hole shadow, respectively, without and with a surrounding geometrically thick and optically thin accretion disk. Bardeen \cite{Bardeen11} studied the Kerr black hole shadow in the bright background. Whereas Cunningham and Bardeen \cite{CT}, investigated how the emission spectrum would change gravitational redshift and lensing, and studied the shadow in the presence of surrounding accretion disk. Several authors \cite{Bardeen11, Takahashi:2004xh,Beckwith:2004ae} have already calculated the photon rings around the Kerr black hole.  The black hole shadow theory has been evolved over the past decades, and the flurry of activities in analytical investigations, observational studies and numerical simulations of shadows and photon rings for modified theories black holes have been elicited \cite{shen1,Cunha1:2016wzk}. The photon rings for compact objects described by the quasi-Kerr metric with characteristic quadrupole moment have also been used as a direct measure of the potential violation of the no-hair theorem \cite{Johannsen:2010ru}. In contrast, a no-horizon ultra-compact object has even number of non-degenerate photon rings \cite{Cunha:2020azh,Cunha:2017qtt}. Moreover, strong gravitational lensing features of no-horizon ultra-compact objects can be vastly different from those of the black holes \cite{Cunha:2017wao,Cunha:2018acu,Cunha:2017qtt,Shaikh:2019itn}. 

Though the concept of the black hole shadow has existed since the '70s, the idea to image it in the Galactic center using the VLBI technique was first presented by Falcke \textit{et al.} in 2000 \cite{Falcke:1999pj}. The Event Horizon Telescope (EHT) collaboration has captured the central compact radio emission region of the nearby elliptical galaxy Messier 87 at $1.3\,$mm wavelength with an unprecedented angular resolution of $~20\,\mu$as \cite{Akiyama:2019cqa,Akiyama:2019fyp,Akiyama:2019eap}. The central flux depression in the horizon-scaled emission image of residing supermassive black hole M87* infers the presence of a black hole shadow with surrounding asymmetric photon ring of diameter $\theta_d=42\pm 3\, \mu $as \cite{Akiyama:2019fyp,Akiyama:2019eap}.  The observed shadow is consistent with the general-relativistic magnetohydrodynamics simulated image of the Kerr black hole. Still, the alternatives to the Kerr black hole also could not be wholly discarded \cite{Akiyama:2019cqa,Akiyama:2019fyp,Akiyama:2019eap,Kumar:2019pjp,Cunha1:2019ikd,Vincent:2020dij}. Moreover, some attentions have been devoted to the shadows and photon rings for exotic alternatives to the black holes, namely wormholes, naked singularities, boson stars, gravastars, and superspinors \cite{Vincent:2020dij,Broderick1:2005xa,Shaikh:2018lcc,Shaikh:2019hbm,Joshi:2020tlq}. These studies concluded that the images of horizonless ultra-compact objects, namely boson stars and Lamy wormhole spacetimes in the presence of accretion disk can be similar to those for the Kerr black holes as long as photon orbits exist around horizonless objects \cite{Vincent:2020dij}. Also, shadows have also been used as a probe of black hole horizon geometry \cite{Cunha1:2018gql}. 

A class of black holes is sourced by the gauge-invariant Lagrangian density $\mathcal{L(F)}$ associated with the nonlinear electrodynamics (NED) field viz. regular black holes.   Bardeen \cite{BardeenReg} using the idea of Sakharov \cite{Sakharov:1966aja} and Gliner \cite{Gliner} proposed the first-ever toy model for the static spherically symmetric regular black hole. 
However, the NED source for the Bardeen black hole arising from the magnetic charge $g$ was obtained much later \cite{AyonBeato:2000zs}. This inspired the construction of several other regular black holes solutions involving the varieties of NED fields \cite{AyonBeato1:1998ub,Hayward:2005gi,Fan1:2016hvf}, also a review \cite{Ansoldi:2008jw}. Moreover, general relativity coupled with the NED field of appropriate Lagrangian density having correct weak-field limits naturally leads to the regular black hole spacetimes \cite{Bronnikov:2000vy,Bronnikov:2000yz}. Whereas Born-Infeld-type NED black holes, which usually do not have fractional power of $F$  are not free from the central curvature singularity \cite{Kruglov:2017mpj,Hendi:2013dwa}; this is because the corresponding NED fields diverge in the strong-field regime. Rotating regular black holes \cite{Bambi:2013ufa,Azreg-Ainou:2014pra,Ghosh:2014pba,Toshmatov:2014nya,Toshmatov:2017zpr} have also been obtained and studied extensively, which in the large$-r$ limits retrieve the Kerr black hole solution \cite{Kerr:1963ud}. 

Intriguing features of photon motions in no-horizon non-rotating regular spacetimes have already been reported in the literature, yielding some fascinating optical phenomenons \cite{Stuchlik1:2014qja,Schee1:2016mjd}. The presence of unstable circular photon orbits and ghost images in the Bardeen and Ayon-Beato-Garcia no-horizon spacetimes are notable among them \cite{Stuchlik1:2014qja}, however, such possibilities in the no-horizon rotating spacetimes are still unexplored to the best of our knowledge. In Refs. \cite{Abdujabbarov:2016hnw,Amir:2016cen,Kumar:2020yem,Kumar:2019pjp,Li:2013jra} the shadows of rotating regular black hole spacetimes are studied, but the cases of no-horizon spacetimes are not discussed.  Further, the cosmic censorship conjecture (CCC) entails that the black hole singularity is always clothed by the event horizon, thereby forbidding naked-singularities \cite{Penrose:1969pc}. CCC's validity is still an outstanding open problem in gravitational physics, and it remains an unproven conjecture to date. The closed photon ring encompassing the shadow ensures the CCC for the Kerr spacetimes \cite{De}, but this may not hold for other rotating black holes.
Furthermore, contrary to the singular black holes, regular black holes can be turned into no-horizon spacetimes with the absorption of charged test particles with finite angular momentum \cite{Li:2013sea}. This motivates us to investigate the photon motions in no-horizon rotating regular spacetimes and derive the photon ring's analytical expression. We are seeking the astrophysical implications of no-horizon rotating spacetimes. In this paper, we investigate the corresponding bound photon orbits and the photon ring's shape and compare them with those for the Kerr black holes and Kerr naked singularity. The Kerr naked singularity always possesses an open arc shadow. Whereas no-horizon rotating regular spacetimes and Kerr-Newman naked singularities may have closed or open photon ring. Indeed, over a finite parameter space ($a, g$), these no-horizon rotating spacetimes possess unstable bound photon orbits and thereby lead to the closed photon ring structure much similar to that of the rotating black hole. This confirms that closed photon ring structure may not always ensure the presence of the black hole horizon. Similar conclusions exist for the spherically symmetric naked singularities, with or without accretion disk \cite{Shaikh:2018lcc,Shaikh:2019hbm,Joshi:2020tlq}. 

This paper is structured as follows: In Sect.~\ref{sect2}, we give an overview of rotating regular black holes. The photon geodesics and formulation of the necessary technique to investigate the surrounding ring structure are discussed in Sect.~\ref{sect3}. Section~\ref{sect4} is devoted to the construction of the photon ring of three familiar rotating regular spacetimes, and we also examine how these can be distinguished from those for the black holes. We conclude in Sect.~\ref{sect5}  with a summary of the key results.

\section{Regular black holes}\label{sect2}
The regular black holes are solutions of general relativity minimally coupled to NED, which yield alteration to classical black holes and near the center behave like a de-Sitter spacetime. The Einstein--Hilbert action minimally coupled with the NED field reads \cite{Salazar:1987ap,AyonBeato:2000zs}
\begin{equation}
I=\int d^4 x\sqrt{-g}\left(\frac{1}{16\pi}R-\frac{1}{4\pi}\mathcal{L(F)}\right),\label{action1}
\end{equation}
where $R$ is the Ricci scalar, and the NED field Lagrangian density $\mathcal{L(F)}$ is a nonlinear and continuous function of invariant $\mathcal{F}=F^{\mu\nu}F_{\mu\nu}/4$, with $F_{\mu\nu}=\partial_{\mu}A_{\nu}-\partial_{\nu}A_{\mu}$ the Faraday electromagnetic field tensor for the gauge potential $A_{\mu}$. This describes Maxwell theory for weak-fields i.e., $\mathcal{L(F)}\approx \mathcal{F}$ as $\mathcal{F} \to 0$, and $\mathcal{L(F)}$ attains a finite values in the strong-field limit $\mathcal{F}\to\infty$. Varying action (\ref{action1}) leads to the following covariant equations of motion \cite{AyonBeato:2000zs}
\begin{eqnarray}
&&G_{\mu\nu}=T_{\mu\nu}=2\left(\mathcal{L_F}\tensor{F}{_\mu}{^\sigma}F_{\nu\sigma}-g_{\mu\nu}\mathcal{L(F)}\right),~\label{Eq1}\\
&&	\nabla_{\mu}\left(\mathcal{L_F}F^{\mu\nu}\right)=0 \quad \text{and}\quad \nabla_{\mu}(^*F^{\mu\nu})=0,\label{Eq2}
\end{eqnarray}
wherein an asterisk denotes the Hodge dual, $G_{\mu\nu}$ is the Einstein tensor, and $\mathcal{L}_{\mathcal{F}}=\partial \mathcal{L(F)}/{\partial\mathcal{F}}$. Bardeen \cite{BardeenReg} presented the first regular black hole model and later several other black hole solutions based on Bardeen's idea have been obtained \cite{AyonBeato1:1998ub,Hayward:2005gi}. The general static and spherically symmetric black hole spacetime metric can be written in the form
\begin{equation}
ds^2=-\left(1-\frac{2m(r)}{r}\right)dt^2+\left(1-\frac{2m(r)}{r}\right)^{-1}dr^2+r^2(d\theta^2+\sin^2\theta d\phi^2),\label{Eq3}
\end{equation}
where $m(r)$ can be determined by using Eq.~(\ref{Eq3}) and solving Eqs.~(\ref{Eq1}) and (\ref{Eq2}). The Faraday field tensor for a spherically symmetric spacetime can only have non-zero electric field $F_{tr}=-F_{rt}$ and magnetic field $F_{\theta\phi}=-F_{\phi\theta}$. For a purely magnetically charged black hole with magnetic monopole charge $g$ and gauge field potential $A=-g\cos\theta\, d\phi$, Eq.~(\ref{Eq2}) yields $F_{\theta\phi}$=$g\sin\theta$ and the Faraday invariant $\mathcal{F}=g^2/2r^4$ is positive definite. Equation~(\ref{Eq1}) leads to only two independent equations \cite{Fan1:2016hvf}
\begin{eqnarray}
-\frac{2m'(r)}{r^2}+2\mathcal{L(F)}=0,\label{Eq05}\\
-\frac{m''(r)}{r}+2\mathcal{L(F)}-\frac{2g^2}{r^4}\mathcal{L}_{\mathcal{F}}=0.\label{Eq5}
\end{eqnarray}
For regular black holes, the NED Lagrangian density has a form \cite{Fan1:2016hvf}, 
\begin{equation}
\mathcal{L(F)}=\frac{\mu\alpha}{g^3}\frac{(2g^2\mathcal{F})^{\frac{\nu+3}{4}}}{\left(1+(2g^2\mathcal{F})^{\frac{\nu}{4}}\right)^{1+\frac{\mu}{\nu}}},\label{lagDensity}
\end{equation}
with this choice of the Lagrangian density, Eqs.~(\ref{Eq05}) and (\ref{Eq5}) admit the solution \cite{Fan1:2016hvf}
\begin{equation}
m(r)=\frac{\alpha r^{\mu}}{\left(r^{\nu}+g^{\nu}\right)^{\frac{\mu}{\nu}}},\label{massFn}
\end{equation}
where, $\mu,\nu\geq 0$ are dimensionless constants, suitably constrained to ensure the asymptotic flatness, with $\mu=0$ leads to the Schwarzschild black hole \cite{Fan1:2016hvf}. In the asymptotic limit $r\to\infty$, the smooth mass distribution function reads $m(r)\approx \alpha$, which can be identified as the black hole mass parameter $M$, i.e., $\alpha=M$. For the spacetime (\ref{Eq3}) with $m(r)$ in Eq.~(\ref{massFn}), the highly densed central region of black hole will be a de-Sitter like region for $\mu\geq 3$. One can choose $\mu$ and $\nu$ suitably to construct exact spherically symmetric regular black hole solutions \cite{Fan1:2016hvf}.  The Bardeen \cite{BardeenReg} and the Hayward \cite{Hayward:2005gi} black holes are encompassed as special cases, respectively, for $\nu=2, \mu=3$, and $\nu=\mu=3$, with the corresponding mass function in Eq.~(\ref{massFn}). If $g\neq 0$, these black hole solutions develop a de-Sitter core at center ($r\to 0$), thereby avoiding the central curvature singularity. One can generate several regular black holes by appropriate choice of the NED Lagrangian density, e.g., if we consider 
\begin{equation}
\mathcal{L(F)}=\mathcal{F}e^{-s(2g^2\mathcal{F})^{1/4}},\label{NSsource}
\end{equation}
and solve Eq.~(\ref{Eq1}), this leads to the exponential mass distribution function 
\begin{equation}
m(r)=Me^{-g^2/{2Mr}},\label{NSmass}
\end{equation}
where $s=g/{2M}$. It encompasses Schwarzschild black hole as special case when $g=0$, and can be identified as Reissner-Nordstr\"{o}m black hole for $r\gg g^2/{2M}$ \cite{Ghosh:2014pba}.\\
Azreg-A\"{i}nou \cite{Azreg-Ainou:2014pra,Azreg-Ainou:2014aqa} modified the standard Newman$-$Janis algorithm by removing the ambiguity about the complex coordinate transformations. This procedure has been successfully used for generating imperfect fluid rotating solutions and generic rotating regular black hole solutions from spherically symmetric static solutions. The rotating counterpart of the spherically symmetric matrix (\ref{Eq3}) in Boyer-Lindquist coordinates have the Kerr-like form and reads as \cite{Bambi:2013ufa,Ghosh:2014pba,Azreg-Ainou:2014pra,Toshmatov:2014nya,Toshmatov:2017zpr}
\begin{eqnarray}\label{rotmetric}
ds^2 & = & - \left( 1- \frac{2m(r)r}{\Sigma} \right) dt^2  - \frac{4am(r)r}{\Sigma  } \sin^2 \theta\, dt \, d\phi +
\frac{\Sigma}{\Delta}dr^2 + \Sigma\, d \theta^2 \nonumber
\\ & &+  \left[r^2+ a^2 +
\frac{2m(r) r a^2 }{\Sigma} \sin^2 \theta
\right] \sin^2 \theta\, d\phi^2,
\end{eqnarray}
with 
\begin{equation}
\Sigma = r^2 + a^2 \cos^2\theta,\qquad \Delta=r^2 + a^2 - 2m(r)r,
\end{equation}
where $a$ is the black hole spin parameter. Thus, in the Eq.~(\ref{rotmetric}), one can replace $m(r)$ as defined in Eqs.~(\ref{massFn}) or (\ref{NSmass}) to get corresponding rotating regular black hole spacetimes, whereas for $m(r)=M$ it describes the Kerr spacetime. The rotating regular metric Eq.~(\ref{rotmetric}) is geodesically incomplete for $r\geq 0$ \cite{Lamy:2018zvj} and globally regular as scalar invariants like Ricci scalar and Riemann scalar are finite everywhere including $r=0$ \cite{Ghosh:2014pba, Bambi:2013ufa}. The horizons of these spacetimes are determined by the real positive roots of 
\begin{equation}
\Delta(r)=r^2+a^2-2m(r)r=0,\label{horizon}
\end{equation}
where $m(r)$ is defined by Eqs.~(\ref{massFn}) or (\ref{NSmass}). It turns out that there exist a set of black hole parameters values such that Eq.~(\ref{horizon}) admits two positive roots, viz., $r_-$ and $r_+$, respectively, corresponding to inner (Cauchy) and outer (event) horizons \cite{Ghosh:2014pba,Bambi:2013ufa}. Indeed, elementary analysis of Eq.~(\ref{horizon}) suggests a critical value $g=g_E$, corresponding to the extremal rotating regular black hole  i.e., $r_-=r_+\equiv r_E$ \cite{ Ghosh:2014pba,Amir:2015pja}. On the other hand, $\Delta(r)=0$ has no physical roots for $g>g_E$, and two distinct positive roots $r_{\pm}$ for $g<g_E$. These two cases describe, respectively, a rotating regular spacetime with no-horizon and regular non-extremal black holes. For instance, for $a=0.30M$, the critical value of $g_E$, for rotating Bardeen spacetime reads $g_E=0.713153M$, while the value of $g_E$ for rotating Hayward spacetimes is $g_E=1.0046M$ and that for rotating nonsingular spacetime is $g_E=1.12081M$. We can say that the critical value $g_E$ is separating the parameter space for rotating regular black holes and no-horizon rotating regular spacetimes. One can also obtain the extremal value of $a=a_E$ for a fixed $g$, such that degenerate horizons exists. 

\section{Geodesics and unstable photon orbits of rotating spacetimes}\label{sect3}
The photon geodesics around compact astrophysical objects are crucially important both from phenomenological and observational perspectives, as they carry the information about the physical processes in the strong-gravity regime \cite{Bardeen:1972fi,Falcke:1999pj}. To study the possible photon ring structure in the rotating regular spacetimes, we use the Hamilton-Jacobi method for the photon geodesics \cite{Carter:1968rr}
\begin{eqnarray}
\label{HmaJam}
\frac{\partial S}{\partial \tau} = -\frac{1}{2}g^{\alpha\beta}\frac{\partial S}{\partial x^\alpha}\frac{\partial S}{\partial x^\beta},
\end{eqnarray}
where $\tau$ is the affine parameter along the geodesics and $S=S(\tau, x^{\alpha})$ is the Jacobi action. Metric  (\ref{rotmetric}) is stationary and axisymmetric, which yield a set of corresponding Killing vectors $\chi^{\mu}_{(t)}$ and $\chi^{\mu}_{(\phi)}$. The projection of these Killing vectors along the photon four-momenta ($p^{\mu}$) entails the integrals of motion \cite{Chandrasekhar:1992}
\begin{equation}
\mathcal{E}=-\chi^{\mu}_{(t)}p_{\mu},\;\;\;\;\;\;  \mathcal{L}=\chi^{\mu}_{(\phi)}p_{\mu},\label{action2}
\end{equation} 
where $\mathcal{E}$ is energy and $\cal{L}$ is the magnitude of axial angular momentum of photon. For the separable solution of the Eq.~(\ref{HmaJam}), the action must have the form  \cite{Carter:1968rr}
\begin{eqnarray}
S=-{\cal E} t +{\cal L} \phi +S_r(r)+S_\theta(\theta) \label{action},
\end{eqnarray}
wherein $S_r(r)$ and $S_{\theta}(\theta)$, respectively, are functions only of the $r$ and $\theta$ coordinates.
The geodesics equations of motion are fully integrable due to the existence of four independent constants of motion associated with two Killing vectors and two irreducible second-rank Killing tensors. Using Eqs.~(\ref{HmaJam}) and (\ref{action}) the complete equations of motion are obtained as follows \cite{Carter:1968rr,Chandrasekhar:1992}
\begin{align}
\Sigma \frac{dt}{d\tau}=&\frac{r^2+a^2}{r^2-2m(r)r+a^2}\left({\cal E}(r^2+a^2)-a{\cal L}\right)  -a(a{\cal E}\sin^2\theta-{\mathcal {L}})\ ,\label{tuch}\\
\Sigma \frac{dr}{d\tau}=&\pm\sqrt{\mathcal{V}_r(r)}\ ,\label{r}\\
\Sigma \frac{d\theta}{d\tau}=&\pm\sqrt{\mathcal{V}_{\theta}(\theta)}\ ,\label{th}\\
\Sigma \frac{d\phi}{d\tau}=&\frac{a}{r^2-2m(r)r+a^2}\left({\cal E}(r^2+a^2)-a{\cal L}\right)-\left(a{\cal E}-\frac{{\cal L}}{\sin^2\theta}\right)\ ,\label{phiuch}
\end{align}
where 
$\mathcal{V}_r(r)\geq 0$ and $\mathcal{V}_{\theta}(\theta)\geq 0$ are related to the effective potentials for radial and azimuthal motions, and, respectively, have the following form  
\begin{eqnarray}\label{06}
\mathcal{V}_r(r)&=&\left((r^2+a^2){\cal E}-a{\cal L}\right)^2-(r^2+a^2-2m(r)r) ({\cal K}+(a{\cal E}-{\cal L})^2),\quad \\ 
\mathcal{V}_{\theta}(\theta)&=&{\cal K}-\left(\frac{{\cal L}^2}{\sin^2\theta}-a^2 {\cal E}^2\right)\cos^2\theta,\label{theta0}
\end{eqnarray}
where constant $\mathcal{K}$ is the Carter integral of motion and indeed plays a crucial role in characterizing the possible latitudinal motion. For $\mathcal{K}=0$, $\theta$-motion get suppressed, and in turn, all photon orbits are restricted only to a plane ($\theta=\pi/2$), yielding unstable circular orbits at the equatorial plane \cite{Chandrasekhar:1992}. For $\mathcal{K}>0$, the photons undergo non-planer motion and form spherical orbits at a constant Boyer-Lindquist radius. Orbits with $\mathcal{K}<0$ are irrelevant for the black hole observations \cite{Chandrasekhar:1992}. On the other hand, $\mathcal{V}_r(r)$ determines the radial motion. In the black hole spacetimes, the radially infalling photons with energy greater than the potential barrier's height get trapped inside the black hole horizon and account for the dark region in the shadow. However, in the no-horizon regular spacetimes, these photons experience repulsive potential at small $r$ and eventually get reflected and then can escape to a faraway observer. Thereby these photons also play an essential role in the optical appearance. Besides the unstable photon orbits, the no-horizon spacetimes also possess stable photon orbits, as the radial potential has both local maxima and minima. Whereas, photons with energy identically same as the height of potential barrier follow the unstable orbits, which experience continuum turning points in their trajectories, and are characterized by the condition \cite{Chandrasekhar:1992} 
\begin{equation}
\left.\mathcal{V}_r\right|_{(r=r_p)}=\left.\frac{\partial \mathcal{V}_r}{\partial r}\right|_{(r=r_p)}=0;\label{vr} 
\end{equation}
whereas instability of orbits obeys condition $$\left.\frac{\partial^2 \mathcal{V}_r}{\partial r^2}\right|_{(r=r_p)}\geq 0,$$
where $r_p$ is Boyer-Lindquist radius of the photon orbit. Unlike for non-rotating black holes, a constant-$r$ motion does not necessarily imply a constant-$\theta$ motion for rotating black holes. To minimize the number of constants of motion, we define the two dimensionless impact parameters $\eta\equiv\mathcal{K}/\mathcal{E}^2$ and $\xi\equiv\mathcal{L}/\mathcal{E}$ \cite{Chandrasekhar:1992}.
Solving Eq.~(\ref{vr}), the critical values of impact parameters for unstable photon orbits read \cite{Abdujabbarov:2016hnw,Tsukamoto:2017fxq}
\begin{align}
\xi_c&=\frac{[a^2 - 3 r_p^2] m(r_p) + r_p [a^2 + r_p^2] [1 + m'(r_p)]}{a [m(r_p) + r_p [-1 + m'(r_p)]]},\nonumber\\
\eta_c&= -\frac{r_p^3 \Big[r_p^3 + 9 r_p m(r_p)^2 + r_pm'(r_p)  [4 a^2 + 2r_p^2+r_p^2 m'(r_p)]-  2 m(r_p) [2 a^2 + 3 r_p^2 + 3 r_p^2 m'(r_p)]\Big]}{a^2 \Big[m(r_p) + r_p [-1 + m'(r_p)]\Big]^2},\label{impactparameter}
\end{align}
which for the Kerr black hole ($m(r)=M$) reduces to \cite{Chandrasekhar:1992}
\begin{align}
\xi_c&=\frac{r_p^2(r_p-3M)+a^2(r_p+M)}{a(M-r_p)},\nonumber\\
\eta_c&=\frac{r_p^3\left(4Ma^2-r_p(r_p-3M)^2\right)}{a^2(M-r_p)^2}.\label{impactparameterkerr}
\end{align}
Likewise $\mathcal{E}$ and $\mathcal{L}$, the impact parameters ($\eta,\xi$) are constant along the geodesics such that depending on their values photons may get scatter ($\xi>\xi_c$), capture ($\xi<\xi_c$), or move along the unstable orbits ($\xi=\xi_c$). That way, for the black hole spacetimes, photons originating from a distant source with impact parameters smaller than the critical values will spiral in and hit the black hole singularity and fail to reach the asymptotic observer, and thereby account for the dark region of the shadow \cite{Bardeen:1972fi,CT}. Whereas photons with critical impact parameter values define the photon ring, therefore, photon rings are the projection along null geodesics of the circular photon orbit \cite{Bardeen11}. The 2$-$1 mapping from the critical values of impact parameter ($\eta_c,\xi_c$) for the unstable photon orbits to the ring coordinates on the observer's screen  ($\alpha,\beta$) reads as \cite{Chandrasekhar:1992,Gralla:2019drh}
\begin{eqnarray}
&&\alpha=\lim_{r_o\rightarrow\infty}\left(-r_o^2 \sin{\theta_o}\frac{d\phi}{d{r}}\right)=-\xi_c\csc\theta_o\xrightarrow{\theta_o=\pi/2} -\xi_c,\nonumber\\ &&\beta=\lim_{r_o\rightarrow\infty}\left(r_o^2\frac{d\theta}{dr}\right)=\pm\sqrt{\eta_c+a^2\cos^2\theta_o-\xi_c^2\cot^2\theta_o}\xrightarrow{\theta_o=\pi/2}\pm\sqrt{\eta_c}, \label{pt}
\end{eqnarray}
where $\beta-$axis is chosen along the rotational axis of the black hole.  
The observer is considered in the far distant asymptotically flat region ($r_o\to\infty$) at an inclination angle $\theta_o=\pi/2$. The parametric plot of $\beta$ vs. $\alpha$ for varying $r_p$ delineates a closed curve on the observer's image plane, which defines a bright photon ring. Although the complete intensity map in the black hole shadow certainly depends on the considered background light distribution, detailed and complex specific accretion model, and emission processes, however, the ring structure and its shape and size are determined solely by the spacetime geometry and the bound photon orbits \cite{Beckwith:2004ae,Johannsen:2010ru,Johannsen:2015qca,Gralla:2019drh}. Therefore, the detection of the closed photon ring can potentially be a clean measurement of the underlying theory of gravity without many of the complications that come with those messy surroundings. Thus, it is definitively intriguing to think about the photon ring structure for the no-horizon rotating regular spacetimes without considering any specific accretion disk model. Moreover, the photon ring is nearly circular for the Kerr black hole and is displaced off center, whereas its shape and size are further altered if the Kerr hypothesis is violated \cite{Johannsen:2015qca}.
In rotating black hole spacetime, photons can have prograde or retrograde motions, their respective circular orbits radii at the equatorial plane, $r_p^{-}$ and $r_p^{+}$, can be determined by solving $\beta=0$ from Eq.~(\ref{pt}) \cite{Chandrasekhar:1992}. Whereas generic spherical photon orbits, having radii $r_p^{-}\leq r_p\leq r_p^+$, construct a photon region around the rotating black hole, which is determined by Eqs.~(\ref{theta0}) and (\ref{impactparameter}), and given by ($\mathcal{V}_{\theta}(\theta)\geq 0$)
\small{\begin{align}
&4 r_p^2a^2[r_p^2+a^2- 2 r_p m(r_p)]\geq\Big[[m(r_p)-r_p (1 - m'(r_p))]a^2 \sin\theta  - \left[(a^2 - 3 r_p^2) m(r_p) + (a^2 + r_p^2) (1 + m'(r_p))r_p \right]\csc\theta \Big]^2,\label{photonR}
\end{align}}
which can be re-written in the simplified form as follow
\begin{equation}
(4r_p\Delta-\Delta'\Sigma)^2\leq 16a^2r_p^2\Delta\,\sin^2\theta,
\end{equation}
the gravitationally lensed image of this photon region corresponds to the black hole shadow. For a given inclination angle $\theta_o$, only a subset of spherical orbits $r_p^{-}\leq r_p\leq r_p^+$ mapped to the photon ring. However, for $\theta_o=\pi/2$, a full set of orbits get imaged on the sky.
Due to Carter's constant motion, the photon ring is symmetric around the horizontal axis, and the only distortion is along the vertical axis. This is because the vertical coordinate of a point on the photon ring is proportional to the corresponding component of the photon 4-momentum $p_{\theta}$, and photons with a positive and a negative $p_{\theta}$ possess identical conserved quantities $\eta$ and $\xi$. For rotating black holes, the photon ring is distorted from a perfect circle and we define the circularity deviation $\Delta C$ to characterize this distortion \cite{Johannsen:2010ru,Johannsen:2015qca}
\begin{equation}
\Delta C=2\sqrt{\frac{1}{2\pi}\int_0^{2\pi}\left(R(\varphi)-\bar{R}\right)^2d\varphi},
\end{equation}
such that $\Delta C=0$ for non-rotating black holes having circular photon rings. Considering a coordinate system with origin at ($\alpha_C,\beta_C$), any point on the ring has radial coordinate $R(\varphi)$ and angular coordinate $\varphi$. The ring average radius $\bar{R}$ is defined as \cite{Johannsen:2010ru}
\begin{equation}
\bar{R}=\frac{1}{2\pi}\int_{0}^{2\pi} R(\varphi) d\varphi.
\end{equation}
The circular asymmetry is primarily affected by the black hole spin, deviation parameters and the observer's inclination angle. In particular, for the M87* black hole shadow the EHT deduced $\Delta C\leq 0.10$ \cite{Akiyama:2019cqa}.

\subsection{Photons orbits around Kerr black hole}
The horizons of the Kerr black hole are located at Boyer-Lindquist radii $r_{\pm}=M\pm\sqrt{M^2-a^2}$, and $r_{\pm}\equiv r_E=M$ for the extremal case ($a=M$). For Kerr black hole, Eq.~(\ref{photonR}) reduces to
\begin{equation}
\Big[\left(r_p^2(r_p-3M)+a^2(r_p+M)\right)^2-a^4(r_p-M)^2\sin^2\theta\Big]\cot^2\theta\leq -r_p^3\Big[r_p(r_p-3M)^2-4Ma^2\Big]. \label{photonRKerr}
\end{equation}
The prograde and retrograde circular photon orbits radii at the equatorial plane read  \cite{Teo:2003,Gralla:2019drh}
\begin{eqnarray}
r_p^-&=&2M\left[1+ \cos\left(\frac{2}{3}\cos^{-1}\left[-\frac{|a|}{M}\right]\right) \right],\nonumber\\
r_p^+&=&2M\left[1+ \cos\left(\frac{2}{3}\cos^{-1}\left[\frac{|a|}{M}\right]\right) \right],
\end{eqnarray}  
\begin{figure} [t!]
	\begin{center}		
		\begin{tabular}{c c}
			\includegraphics[scale=0.83]{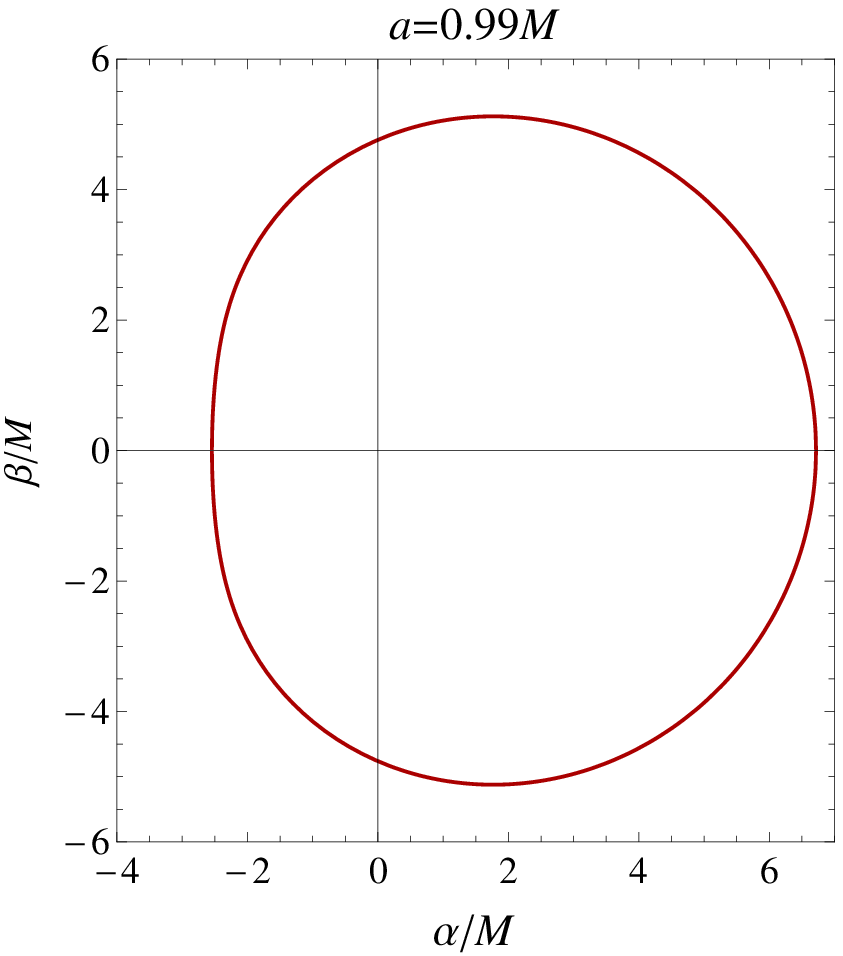} &
			\includegraphics[scale=0.83]{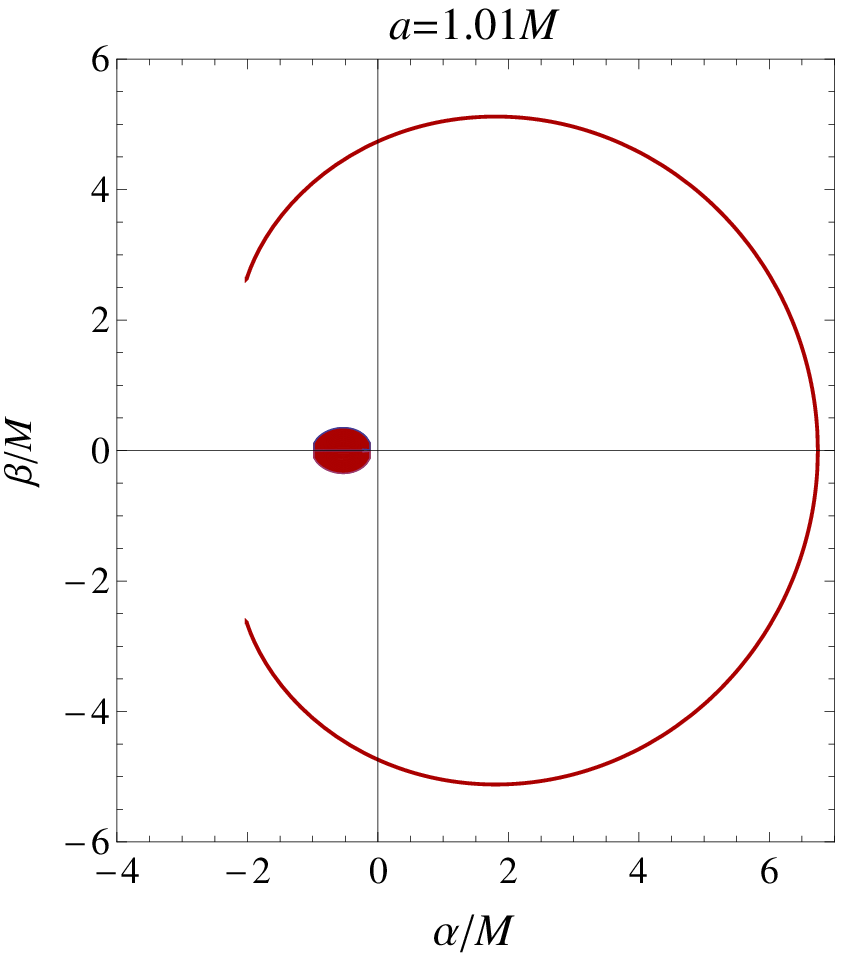} 
		\end{tabular}
		\caption{Photon ring around the Kerr black hole (left) and the Kerr naked singularity (right) for inclination angle $\theta_o=\pi/3$.}\label{KerrNS}
	\end{center}
\end{figure}
\begin{figure}
	\begin{center}
		\includegraphics[scale=0.75]{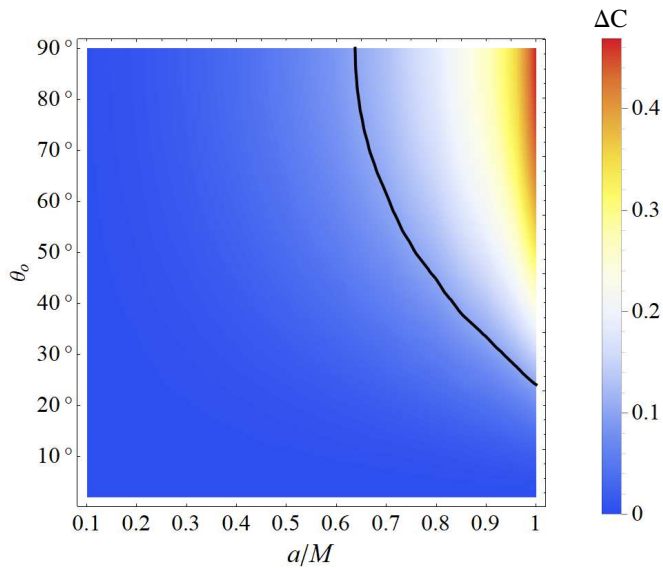}
		\caption{The photon ring circularity deviation observable $\Delta C$ for the Kerr black hole as a function of ($a, \theta_o$). The black solid line is for $\Delta C=0.10$, such that the region on the right-side of the black line is excluded by the measured circularity for the	M87* black hole reported by the EHT, $\Delta C\leq 0.10$.}\label{Km87}
	\end{center}
\end{figure}
and they fall in the range $M\leq r_p^-\leq 3M$ and  $3M\leq r_p^+\leq 4M$. Indeed, in the limit of zero rotation ($a=0$), these two circular orbits coincide into a single orbit, and the photon region in Eq.~(\ref{photonRKerr}) degenerate into a photon sphere of constant radius, e.g., for the Schwarzschild black hole $r_p=3M$. The $r_p^-\leq r_p^+$ can be attributed to the Lense-Thirring effect \cite{Johannsen:2010ru} and due to spacetime dragging photons revolve around the black hole with respect to the static observer at spatial infinity. For extremal Kerr black hole ($a=M$), the prograde orbits and event horizon have the same radii $r_p^-=r_E=M$, however, they are still separated by a finite proper distance \cite{Bardeen:1972fi}. Photon ring circularity deviation depends on the observer's inclination angle $\theta_o$ (cf. Eq.~(\ref{pt})), such that it is maximum for $\theta_o=\pi/2$ and further increases with $a$. However, for $\theta_o=0, \pi$, the apparent photon ring is perfectly circular for arbitrary values of $a$. Photons orbiting up to the polar plane with radii $r_p^0$ that is determined by  $\xi_c=0$ construct this circular photon ring. Thus, Kerr black holes ($a\leq M$) always have prograde and retrograde unstable photons orbits and form a closed photon ring or shadow silhouette \cite{Bardeen:1972fi}. Indeed, in the Kerr spacetimes, the ring closedness is equivalent to the CCC ~\cite{De}. Furthermore, for $a>M$, suffice to ascertain a Kerr naked singularity, prograde photons on the equatorial plane and a few neighboring spherical photon orbits spiral in and eventually end up in the central singularity, thereby the retrograde photon orbits, relatively less affected by the black hole frame-dragging, cause an open arc ring structure with a tiny segment missing near the left endpoint (cf. Fig.~\ref{KerrNS}) \cite{Charbulak:2018wzb,Wilkins:1972,Hioki:2009na}. In the Kerr naked singularity spacetime, orbits with $r_p<0$ are also crucial for $\theta_o\neq \pi/2$, because they escape into the another spatial infinity ($r\to -\infty$) and never reaches the faraway observer ($r_o\to \infty$) \cite{Hioki:2009na}. Indeed, these photon orbits account for a dark central spot in the naked singularity shadows (cf. Fig.~\ref{KerrNS}).

Figure~\ref{KerrNS} shows that the photon ring structure for the Kerr naked singularity is markedly different from that of a Kerr black hole. For a Kerr black hole, the shape of this photon ring is nearly circular unless the black hole spins very rapidly. On the other hand, Reissner-Nordstr\"{o}m black hole spacetime exhibits an intriguing feature that the photon ring can be closed even if CCC is violated ($Q>M$), in fact, naked singularities possess closed photon ring for $Q<\sqrt{9/8}M$ \cite{De,Zakharov:2014lqa}. This is because of the presence of outer unstable circular photon orbits in the Reissner-Nordstr\"{o}m naked singularity spacetimes. The  Kerr black hole's photon ring circularity deviation $\Delta C$ is shown as a function of ($a, \theta_o$) in Fig.~\ref{Km87}. Clearly $\Delta C$ vanishes for $\theta_o=0$  and increases with $\theta_o$. The circularity deviation bound for the M87* black hole shadow, $\Delta C\leq 10 \%$, places a constraint on the inclination angle and the Kerr black hole spin parameter.

Since the occurrence of the naked singularity leads to the violation of the CCC, however, its understanding will help to put CCC in concrete mathematical form. Indeed, some spherically symmetric naked singularity models, Janis-Newman-Winicour \cite{Shaikh:2019hbm} and Joshi-Malafarina-Narayan naked singularities \cite{Shaikh:2018lcc}, for some particular values of parameters, possess photon spheres and cast dark shadows that are much similar in nature to the Schwarzschild black hole shadow. However, in the absence of photon spheres, the images of naked singularities significantly differ from those of black holes \cite{Shaikh:2019hbm}. This motivates us to investigate the other possibilities of the no-horizon cases of rotating spacetimes, resembling the black hole shadows. 

\section{Photon rings of no-horizon rotating regular spacetime}\label{sect4}
In this section, we use the above technique to determine how spherical photon orbits get modified because of the deviation parameter $g$. We compare photon ring produced by the Kerr black hole and Kerr naked singularity with those produced by the three well-known rotating regular no-horizon spacetimes, described by viz. Bardeen, Hayward and nonsingular. 

\subsection{Bardeen spacetime}
The rotating Bardeen black hole belongs to the prototype non-Kerr family and described by metric (\ref{rotmetric}) with the mass function 
\begin{equation}
m(r)=M\left(\frac{r^2}{r^2 + g^2}\right)^{3/2}, \label{Bardeenmass}
\end{equation}
such that in the limit $g\to 0$, it retrieve the Kerr black hole. The rotating Bardeen black hole shadows are smaller and more distorted than the Kerr black hole shadows \cite{Abdujabbarov:2016hnw,Li:2013jra,Tsukamoto:2014tja,Kumar:2020yem}.\\
\begin{figure*}[b!]
	\begin{tabular}{c c}
		\includegraphics[scale=0.8]{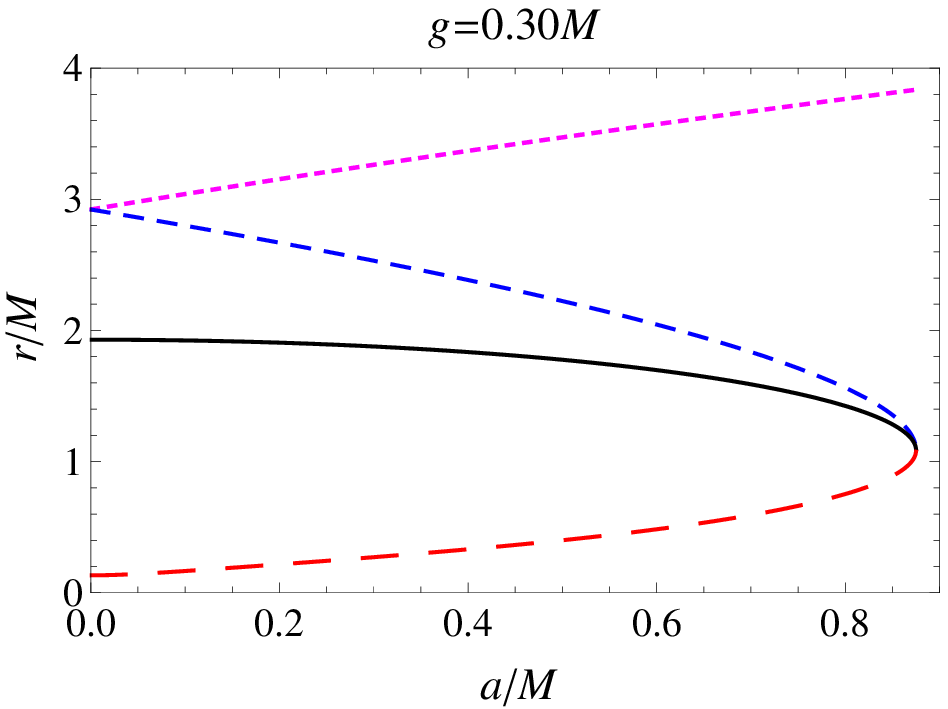}&
		\includegraphics[scale=0.8]{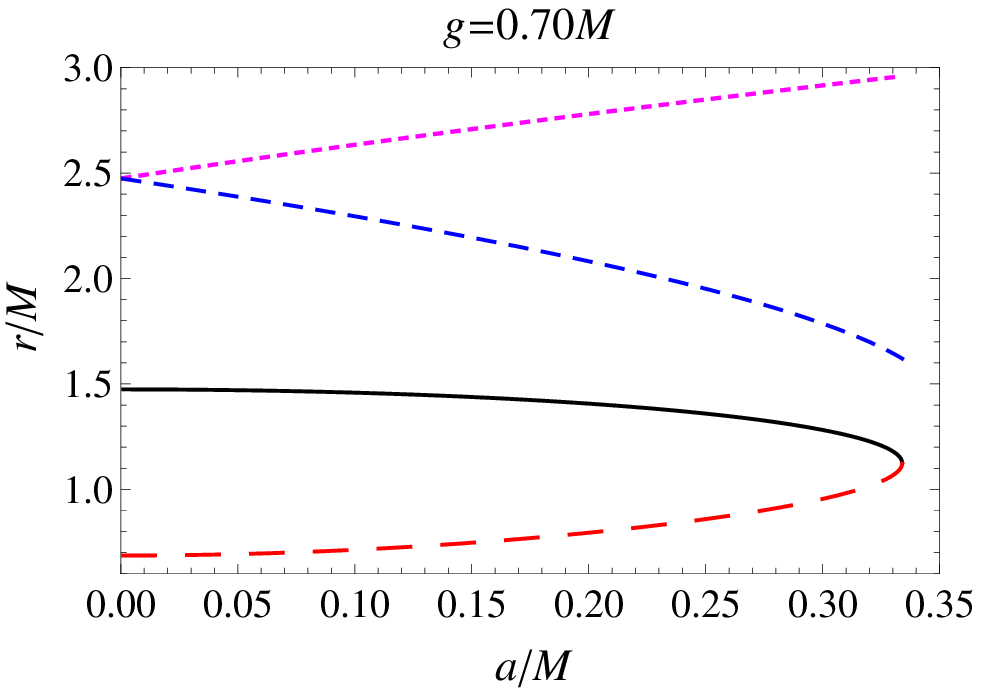}\\
	\end{tabular}
	\caption{Radii of Cauchy horizon $r_-$ (\textit{Red dashed curve}), event horizon (\textit{Black solid curve}),  prograde orbit $r^-_p$ (\textit{Blue small dashed curve}), and retrograde orbit $r^+_p$ (\textit{Magenta dotted curve}) with varying $a$ for rotating Bardeen black hole.}\label{Bardeenradii}
\end{figure*}
\begin{table}
	\centering
	\begin{tabular}{ |p{1.5cm}|p{1.8cm}|p{1.8cm}|p{1.8cm}|p{1.8cm}|p{2.9cm}| }
		\hline
		$g/M$ &  $a_{E}/M$ & $r_{E}/M$ & $r^-_p/M$ &  $r^+_p/M$ & $\delta=(r^-_p-r_{E})/M$  \\
		\hline\hline
		0.0 &  1 &  1 & 1 &  4  &  0 \\
		\hline
		0.10 &  0.985177 &  1.01406 & 1.01406 & 3.98135 &  0 \\
		\hline
		0.20 &  0.942439 &  1.04762 & 1.04762 & 3.92633 &  0 \\
		\hline
		0.30 &  0.875075 &  1.08604 & 1.08604 & 3.83598 &  0 \\
		\hline
		0.40 &  0.785157 &  1.11879 & 1.11879 & 3.70880 &  0 \\
		\hline
		0.50 &  0.671965 &  1.13979 & 1.13979 & 3.53828 &  0 \\
		\hline
		0.60 &  0.529499 &  1.14404 & 1.23056 & 3.30719 &  0.08651 \\
		\hline
		0.70 &  0.334007 &  1.12449 & 1.62061 & 2.96033 &  0.49612 \\
		\hline
		0.769 &  0.035112 &  1.08923 & 2.22920 & 2.37297 & 1.13997 \\
		\hline 
		0.76980 &  0.0 &  1.08866 & 2.30118 & 2.30118 & 1.21252 \\
		\hline 
	\end{tabular}
	\caption{Table summarizing the values of extremal horizon radius $r_{E}$, prograde and retrograde photon orbit radii $r^-_p$, $r^+_p$ for rotating Bardeen black hole. For extremal black holes photon region size deceases with increasing $g$. }\label{BardeenTable}
\end{table}
\begin{figure*}
\begin{center}
\includegraphics[scale=0.85]{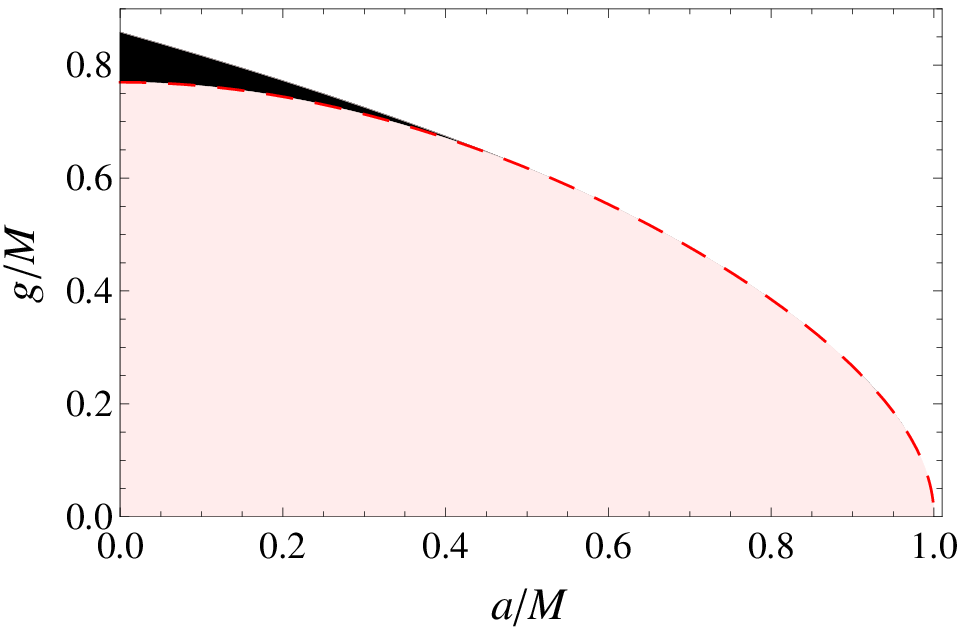}
	\caption{Parameter plane ($a, g$) for the rotating Bardeen spacetime. The red dashed line separates the black hole spacetimes from the no-horizon spacetimes. The no-horizon spacetimes also admit closed photon ring for parameters ($a, g$) within the black shaded region.}\label{bar}
\end{center}
	\begin{tabular}{c c}
		\includegraphics[scale=0.85]{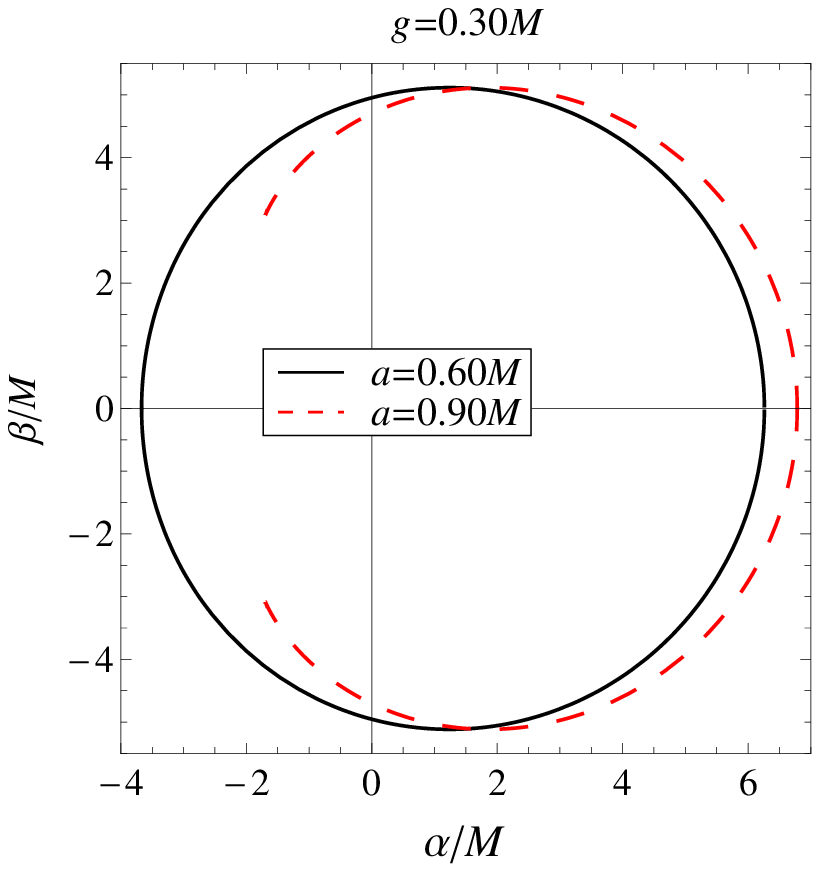}&
		\includegraphics[scale=0.85]{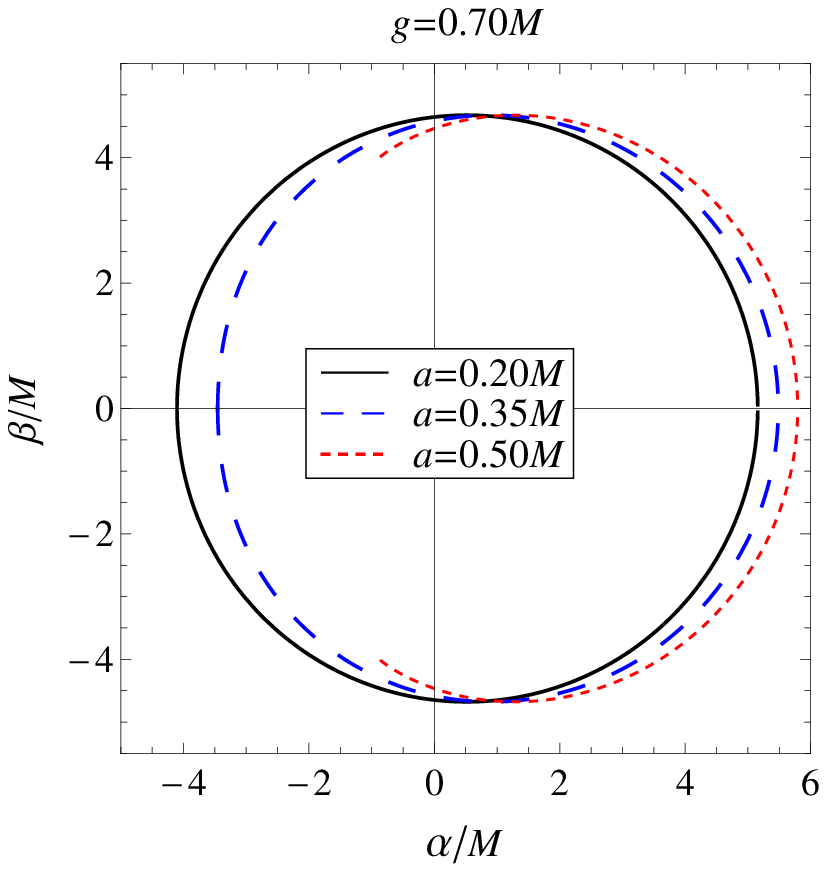}\\
	\end{tabular}
	\caption{Comparison of photon rings for rotating Bardeen spacetime. Black solid curves correspond to those for black holes, whereas dashed and dotted curves are for the no-horizon spacetimes.}\label{BardeenShadow}
	\label{Bardeen}
\end{figure*}
\begin{figure*}
	\begin{center}	
		\includegraphics[scale=0.75]{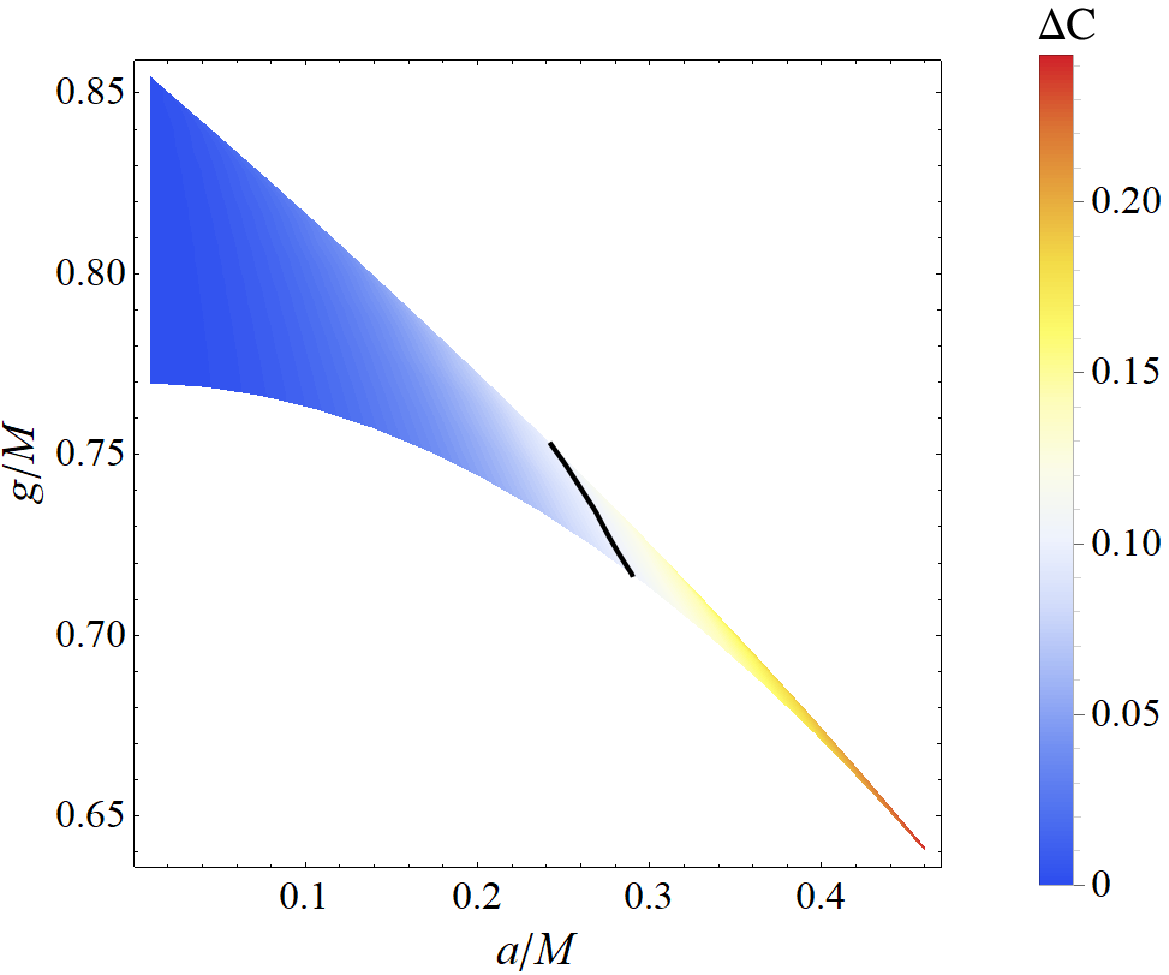}
		\caption{The photon ring circularity deviation observable $\Delta C$ for the no-horizon rotating Bardeen spacetimes as a function of ($a, g$). The solid black line is for $\Delta C=0.10$, such that the region on the right-side of the black line is excluded by the measured circularity for theM87* black hole reported by the EHT, $\Delta C\leq 0.10$.}\label{Bm87}
	\end{center}
\end{figure*}
The variation of horizon radii $r_{\pm}$ and unstable circular photon orbit radii $r_p^{\pm}$ with spin parameter $a$ for rotating Bardeen black holes are depicted in Fig. \ref{Bardeenradii}. The photon region around the black hole grows with $a$ but considerably decreases with $g$. For small $g$ ($g\leq 0.572M$) and increasing $a$, prograde orbits approach the event horizon and both radii eventually coincide for $a=a_E$. However, for sufficiently large values of $g$ ($> 0.572M$) there exists $a=a_E$ corresponding to the extremal black hole, for which the prograde orbits $r_p^-$ do not merge with the extremal horizon $r_E$ (cf. Fig. \ref{Bardeenradii}), i.e., $\delta=r_p^-- r_E\neq 0$. Table \ref{BardeenTable} summarizes the extremal horizon radius $r_{E}$ and circular orbit radii $r_p^{\pm}$ for various extremal black hole configurations, such that $a_E$ decreases with increasing $g$. For non-rotating Bardeen black hole with $g=g_E=0.7698M$, only one unstable circular photon orbit exist outside the event horizon at $r_p=2.30118M$ whereas extremal horizon radii is $r_E=1.0886M$. However, in the no-horizon spacetimes, two circular photon orbits, inner being stable and outer being unstable, can exist in the interval $g_E<g<g_c=0.858650M$, and for $g=g_c$  both coincide at $r_p=1.717M$ and no circular photon orbit can exist for $g>g_c$ \cite{Stuchlik1:2014qja}. This is because of the fact that for $a=0$ and $g=g_E$, circular photon orbits do not merge with the black hole horizon (cf. Table~\ref{BardeenTable}). For some particular values of parameters ($a,g$), shown as the black shaded region in  Fig.~\ref{bar}, unstable circular photon orbits, especially prograde, exist in the no-horizon rotating Bardeen spacetimes as well, which is not true for the Kerr naked singularities. For a black hole having anti-clockwise rotation, the prograde and retrograde orbits, respectively, construct the left and right portion of the photon ring. Therefore, the presence of both orbits is crucial for the closed photon ring structure, such that in their absence, retrograde orbits solely outline an open arc structure. Photon rings for rotating Bardeen spacetimes for different values of $a$ and $g$ are depicted in Fig. \ref{BardeenShadow} and are compared with the corresponding no-horizon rotating spacetime counterparts. For small values of $g$ ($\leq 0.572M$), the photon ring of no-horizon rotating Bardeen spacetimes ($a>a_E$) are always open arc with a tiny missing segment near the left edge and distinct from that of rotating Bardeen black hole ($a<a_E$), which have closed shape. Interestingly, photon rings of no-horizon rotating Bardeen spacetimes are similar to that of Kerr naked singularity, except that now they do not have a central dark spot due to photon orbits from $r<0$. However, when $g$ is large ($g> 0.572M$), the no-horizon rotating Bardeen spacetimes may possess prograde orbits, and thus the photon ring may still have a closed shape (cf. Fig. \ref{BardeenShadow}). 

In Fig.~\ref{BardeenShadow}, we have shown that photon ring of no-horizon spacetime ($g=0.70M, a=0.35M>a_E$) is indeed a closed curve with circularity deviation observable $\Delta C=0.1588$. For $a\gg a_E$, the prograde orbits disappear and, the photon ring again reduces into an open arc (cf. Figs.~\ref{bar} and \ref{BardeenShadow}). Thus, the no-horizon rotating Bardeen spacetimes, with some specific values of parameters shown as a black shaded region in Fig.~\ref{bar}, possess closed photon rings that resemble those for the rotating black holes. Therefore, the presence of a closed photon ring does not ensure the existence of the black hole horizon. Figure~\ref{Bm87} shows the circularity deviation observable $\Delta C$ as a function of $a$ and $g$ for the no-horizon rotating Bardeen spacetimes possessing closed photon rings. Clearly, $\Delta C$ depends on the spin $a$ and NED charge $g$, also the M87* bound $\Delta C\leq 0.10$ is satisfied.

\subsection{Charged Hayward spacetime}
Frolov \cite{Frolov:2016pav} obtained the spherically symmetric charged Hayward black hole solutions. Later, Kumar \textit{et al.} \cite{Kumar:2019pjp} derived the rotating counterpart solutions, whose line element is described by metric (\ref{rotmetric}) with the mass function $m(r)$ 
\begin{figure}[b!]
	\begin{tabular}{c c}
		\includegraphics[scale=0.8]{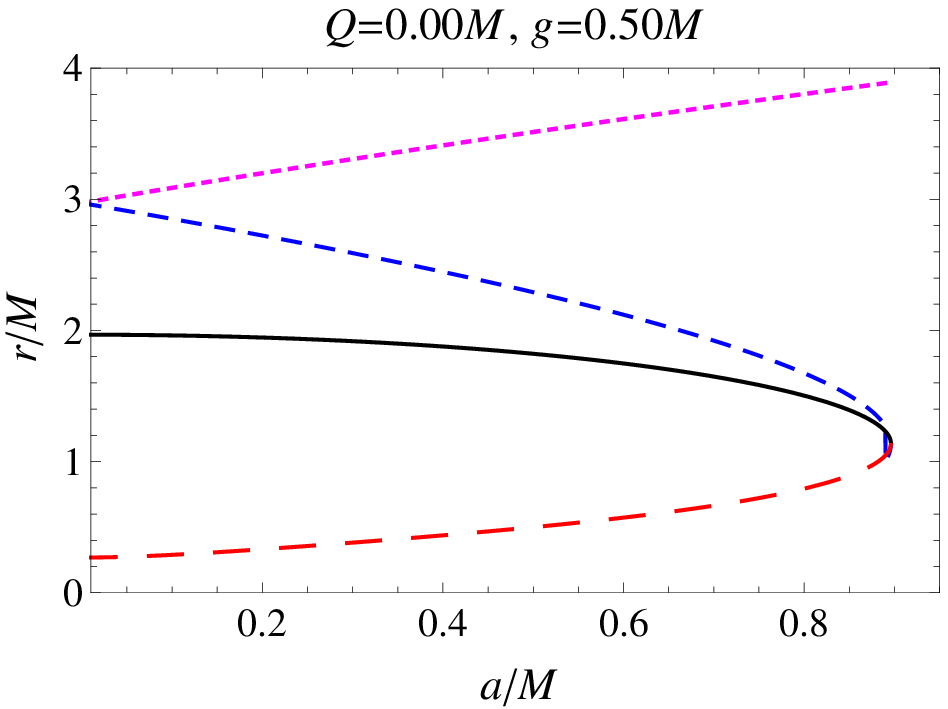}&
		\includegraphics[scale=0.8]{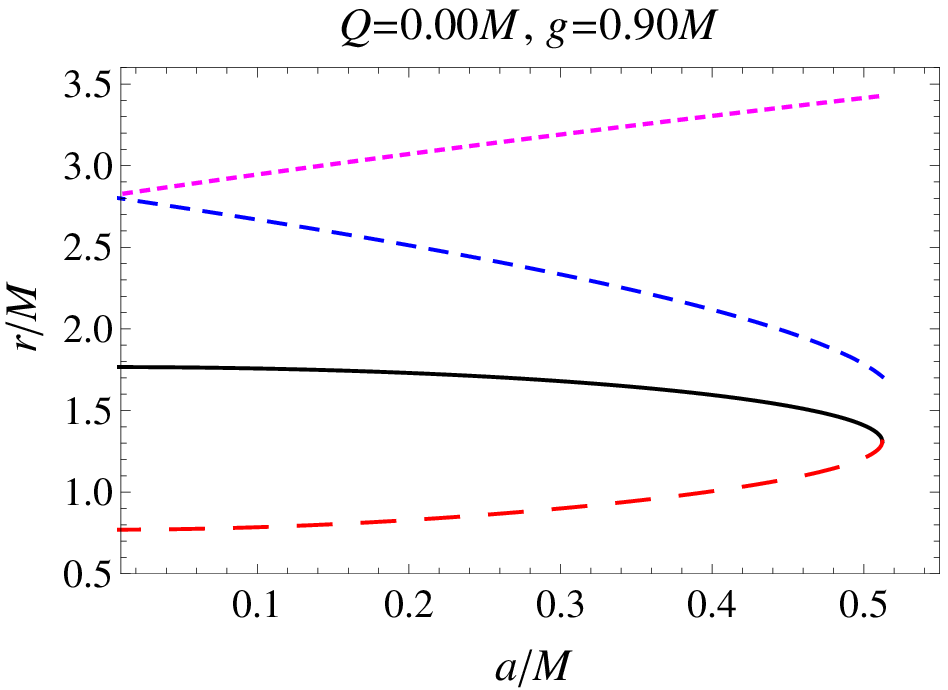}\\
		\includegraphics[scale=0.8]{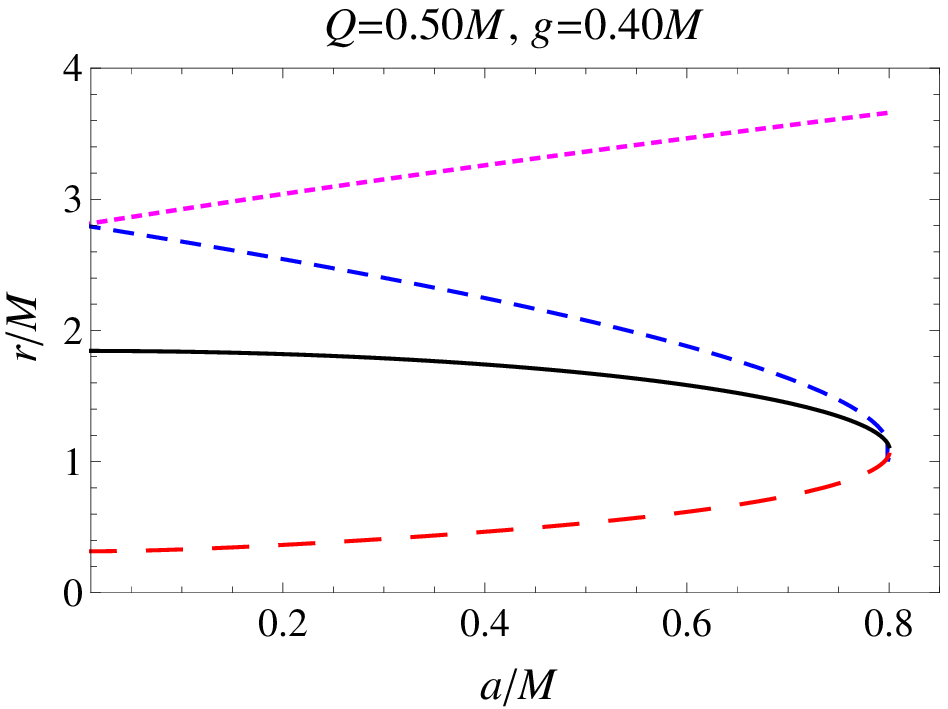}&
		\includegraphics[scale=0.8]{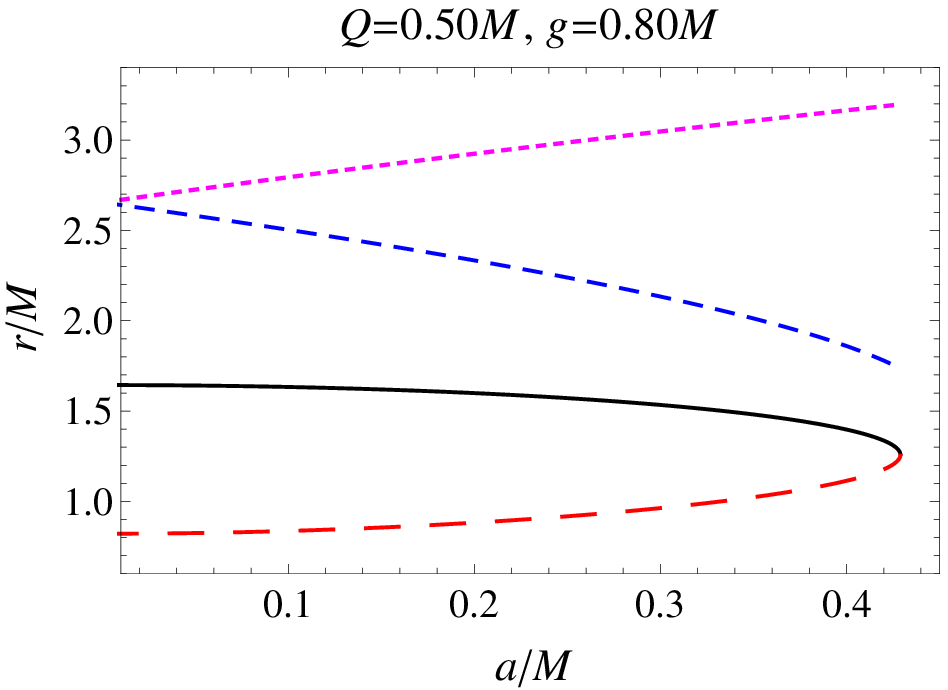}
	\end{tabular}
	\caption{Radii of Cauchy horizon $r_-$ (\textit{Red dashed curve}), event horizon $r_+$ (\textit{Black solid curve}),  prograde orbit $r^-_p$ (\textit{Blue small dashed curve}), and retrograde orbit $r^+_p$ (\textit{Magenta dotted curve}) for rotating charged Hayward black hole.} \label{HayRadius}
\end{figure}
\begin{table*}[th!]
	\centering
	\begin{tabular}{ |p{1.5cm}|p{1.8cm}|p{1.8cm}|p{1.8cm}|p{1.8cm}|p{2.9cm}| }
		
		\hline
		$g/M$ &  $a_{E}/M$ & $r_{E}/M$ & $r^-_p/M$ &  $r^+_p/M$ & $\delta=(r^-_p-r_{E})/M$  \\
		\hline\hline
		0.0 &  1 &  1 & 1 & 4 &  0 \\
		\hline
		0.10 &  0.999002& 1.00198& 1.00198& 3.999&  0.0 \\
		\hline
		0.20 & 0.99215& 1.01501& 1.01501& 3.99212 &  0.0 \\
		\hline
		0.30 & 0.974509& 1.04466& 1.04466& 3.97424 &  0.0 \\
		\hline
		0.40 &  0.942997& 1.08796& 1.08796& 3.9417 &  0.0 \\
		\hline
		0.50 &  0.896106& 1.13802& 1.13802& 3.89193& 0.0 \\
		\hline
		0.60 &  0.832765& 1.18888& 1.18888& 3.82226 &  0.0 \\
		\hline
		0.70 &  0.751193& 1.23642& 1.23642& 3.7285& 0.0 \\
		\hline
		0.80 & 0.647497& 1.27766& 1.42044& 3.60281& 0.142784 \\
		\hline
		0.90 &  0.512192& 1.30988& 1.70778& 3.42781& 0.397897 \\
		\hline
		1.05827 &  0.0&1.33333 &2.65236 &2.65236 & 1.31903 \\
		\hline
	\end{tabular}
	\caption{Table representing the values of extremal horizon radius $r_{E}$, prograde and retrograde photon orbit radii, respectively, $r^-_p$ and $r^+_p$ for Hayward black hole with $Q=0.0$.}\label{HayTable}
	\centering
\begin{tabular}{ |p{1.5cm}|p{1.8cm}|p{1.8cm}|p{1.8cm}|p{1.8cm}|p{2.9cm}| }
	
	\hline
	$g/M$ &  $a_{E}/M$ & $r_{E}/M$ & $r^-_p/M$ &  $r^+_p/M$ & $\delta=(r^-_p-r_{E})/M$  \\
		\hline\hline
		0.0 & 0.866025& 1.0& 1.0& 3.73205& 0.0\\
		\hline
		0.10 & 0.864891& 1.00192& 1.00192& 3.73085 &  0.0 \\
		\hline
		0.20 &0.857094& 1.01453& 1.01453& 3.72258 &  0.0 \\
		\hline
		0.30 & 0.836956& 1.04316& 1.04316& 3.70096 &  0.0 \\
		\hline
		0.40 &  0.800728& 1.08475& 1.08475& 3.66125 &  0.0\\
		\hline
		0.50 & 0.746117& 1.13234& 1.13234& 3.59947 &  0.0 \\
		\hline
		0.60 & 0.670646& 1.17985& 1.17985 &3.51044&  0.0 \\
		\hline
		0.70 & 0.569346& 1.22287& 1.42794& 3.38437& 0.205072 \\
		\hline
		0.80 & 0.428932& 1.25787& 1.73516& 3.19725& 0.47729 \\
		\hline
		0.90 &0.183365& 1.28094& 2.21264& 2.8352& 0.931701 \\
		\hline
		0.921771 & 0.0 & 1.28385& 2.53615& 2.53615& 1.25231\\
		\hline
	\end{tabular}
	\caption{Table representing the values of extremal horizon radius $r_{E}$, prograde and retrograde photon orbit radii, respectively, $r^-_p$ and $r^+_p$ for charged Hayward black hole with $Q=0.50M$.}\label{HayTable1}
\end{table*}

\begin{figure*}[h!]
\begin{center}
	\begin{tabular}{c c}
			\includegraphics[scale=0.8]{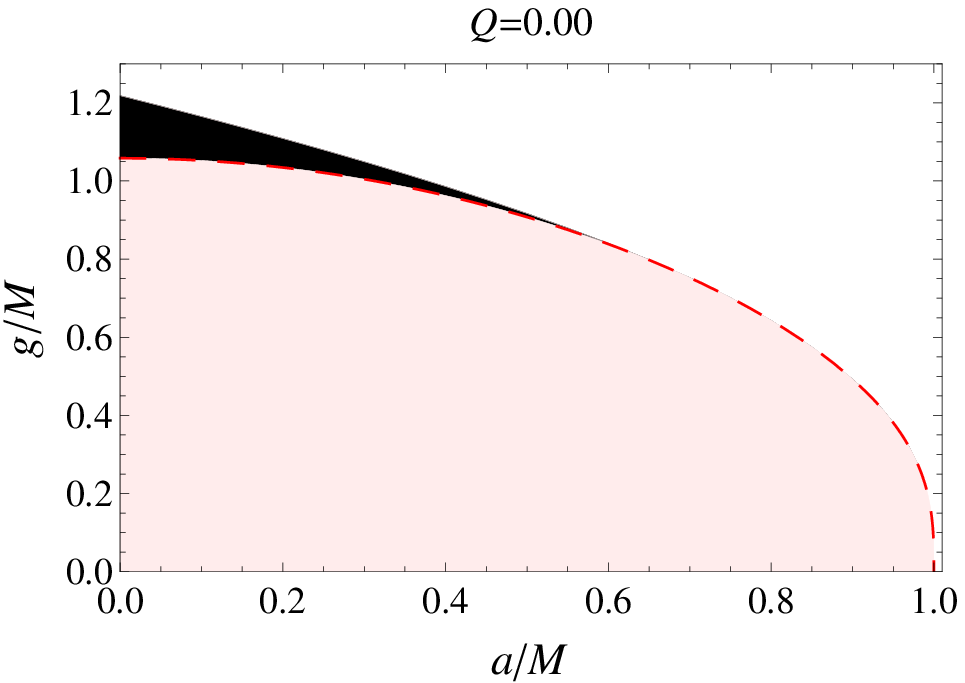} &
			\includegraphics[scale=0.8]{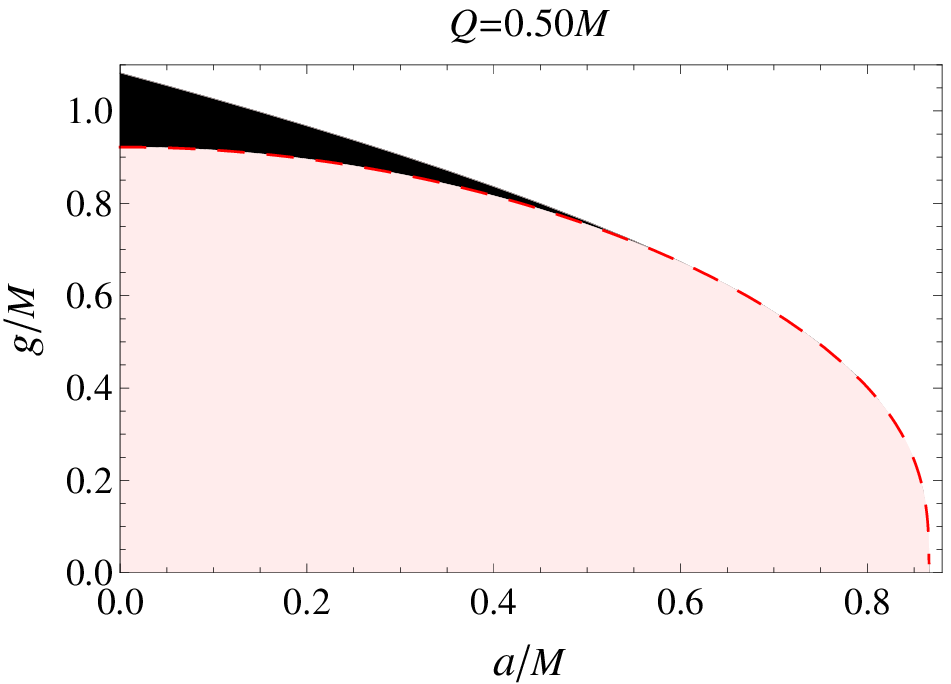}
	\end{tabular}
		\caption{Parameter plane ($a, g$) for the charged rotating Hayward spacetime. The red dashed line separates the black hole spacetimes from the no-horizon spacetimes. The no-horizon spacetimes also admit closed photon rings when parameters ($a, g$) lie in the black shaded region.}\label{CH}
\end{center}
\end{figure*}
\begin{figure*}[h!]
\begin{center}			
	\begin{tabular}{c c}
			\includegraphics[scale=0.75]{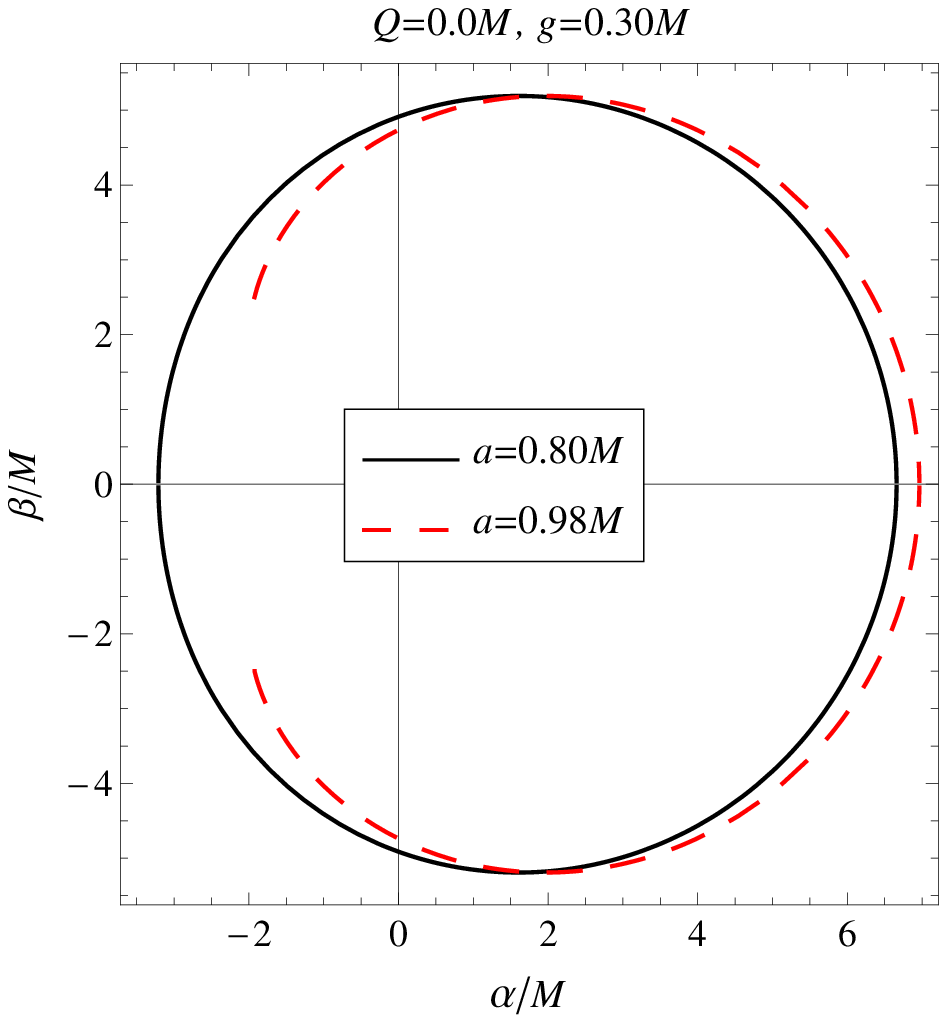}&
			\includegraphics[scale=0.75]{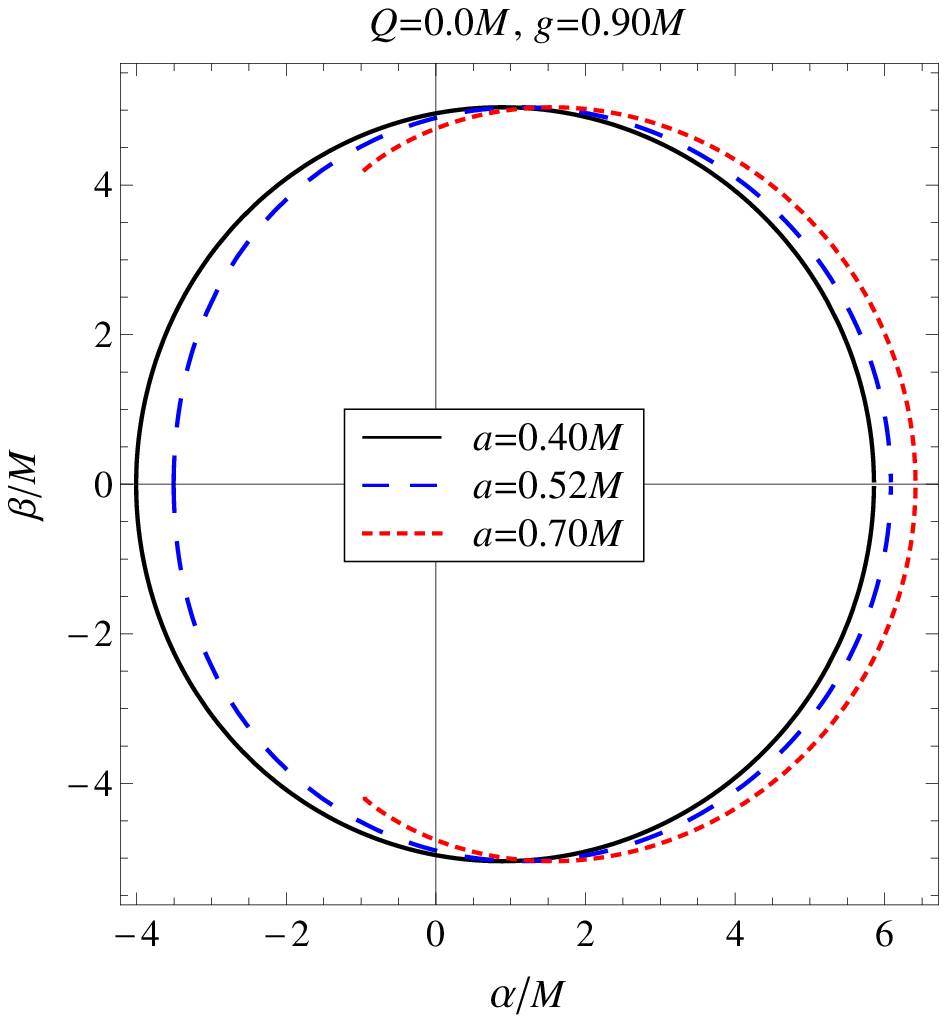}\\
			\includegraphics[scale=0.75]{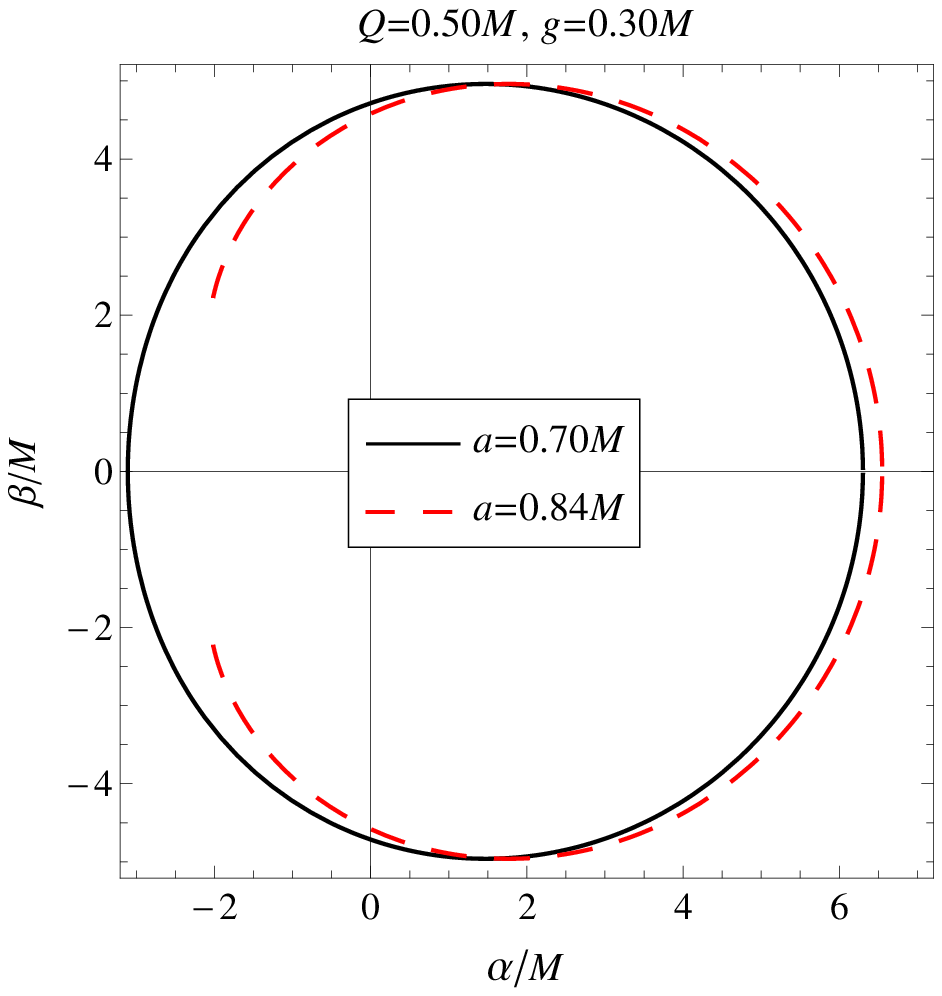}&
			\includegraphics[scale=0.75]{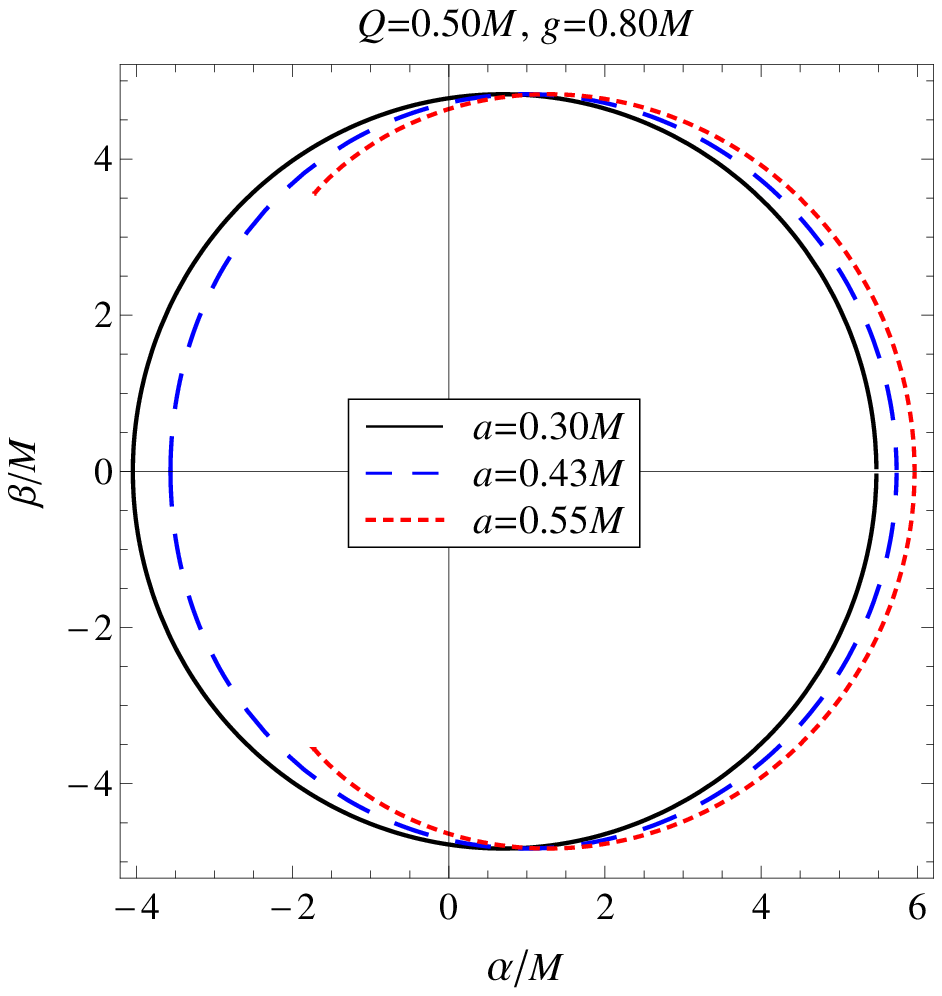}
	\end{tabular}
		\caption{Comparison of photon rings for charged rotating Hayward spacetime. Black solid curves correspond to those for black holes, whereas dashed and dotted curves are for the no-horizon spacetimes.}\label{HayShadow}
\end{center}	
\end{figure*}
\begin{equation}
m(r)=\frac{M\left(2Mr-Q^2\right)r^3}{2Mr^4+g^3\left(2Mr+Q^2\right)},
\end{equation}
where $Q$ is the electric charge. The parameter $g$ is related to the length $\ell$ associated with the region concentrating the central energy density via $g^3=2M\ell^2$, such that modifications in the spacetime metric appear when the curvature scalars become comparable with $\ell^{-2}$ \cite{Hayward:2005gi}. We shall now discuss the spherical photon orbits around the charged rotating Hayward black holes. Abdujabbarov \textit{et al.} \cite{Abdujabbarov:2016hnw} studied the shadows and photon rings cast by rotating regular Hayward black hole ($Q=0$) to infer that the shadow size decreases and become more distorted with $g$. Later, Kumar \textit{et al.} \cite{Kumar:2019pjp} extended this result for the charged rotating Hayward spacetimes ($Q\neq 0$).\\
The behaviour of horizon radii $r_{\pm}$ and unstable circular photon orbit radii $r_p^{\pm}$ are analyzed and depicted in Fig. \ref{HayRadius}. The retrograde photons orbits radii increase while prograde orbits radii decrease with $a$.

In Table \ref{HayTable} and \ref{HayTable1}, we summarize horizon radii $r_{E}$ and unstable circular photon orbit radii $r_p^{\pm}$, for the case of extremal charged Hayward black holes, respectively, for $Q=0$ and $Q=0.5M$. From Tables \ref{HayTable} and \ref{HayTable1}, one can notice that the radii $r_E$ and $r_p^-$ increase with $g$, while $a_E$ decreases with $g$ and so is $r_p^+$. It turns out that for a given $g$ ($g\leq 0.731M$ for $Q=0$ and $g\leq 0.602M$ for $Q=0.50M$), there exists $a=a_E$ corresponding to the extremal black hole, such that the radial coordinate of prograde orbits and extremal horizon radius coincides, i.e., $\delta\equiv r_p^--r_{E}=0$. However, when $g> 0.731M$ for $Q=0$ and $g> 0.602M$ for $Q=0.50M$, they do not coincide, viz. $\delta >0$ and that $\delta$ increases with $g$. A numerical study of $\beta=0$ shows that prograde orbits may prevail for the no-horizon rotating Hayward spacetimes resulting into the alteration of the photon ring structure. The range of parameters ($a, g$) that allows a closed photon ring to occur is shown in Fig.~\ref{CH}. Such as, for the non-rotating cases with $Q=0.0$ and $Q=0.50M$, photon circular orbits exist for $g\leq g_c=1.21834M$ and $g\leq g_c= 1.082293M$, respectively, whereas $g_E=1.05827M$ and $g_E=0.921771M$. For rotating case $a=0.10M$, $g_c=1.164846M$ and $g_E=1.05297M$ for $Q=0.0$, whereas $g_c=1.026494M$ and $g_E=0.901829M$ for $Q=050M$. The photon rings of rotating Hayward spacetimes for different values of parameters ($a, g, Q$) are depicted in Fig.~\ref{HayShadow}. For some particular values of parameters ($a, g$), shown as black shaded region in Fig.~\ref{CH}, the photon ring of no-horizon rotating regular spacetimes have closed structure and resemble those for the rotating black hole (cf. Fig.~\ref{HayShadow}). However, for significantly large violations of extremal bound ($a>a_E$), the ring is again an open curve as shown in Fig.~\ref{HayShadow}. 

\begin{figure*}
\begin{center}
 \begin{tabular}{ c c}	
	\includegraphics[scale=0.7]{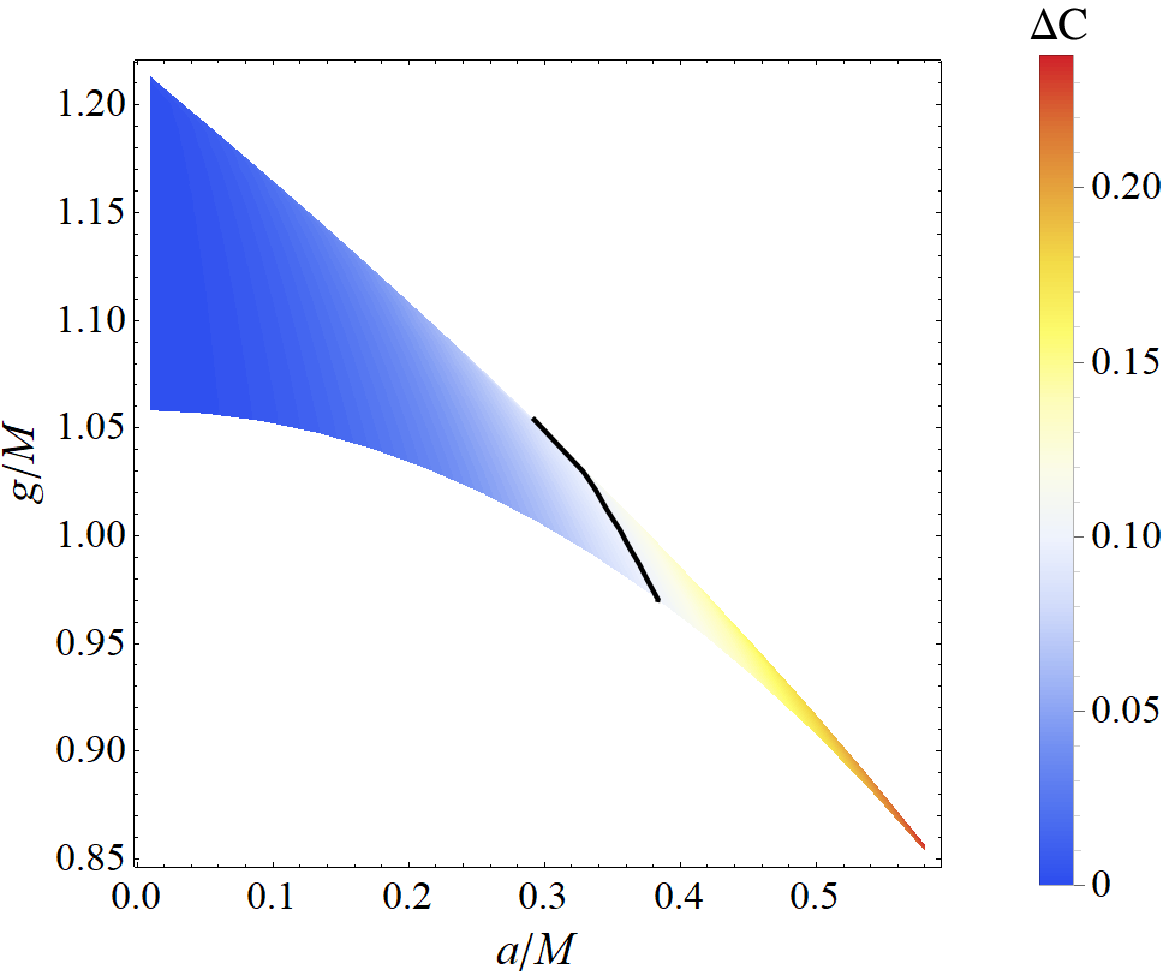}&
	\includegraphics[scale=0.7]{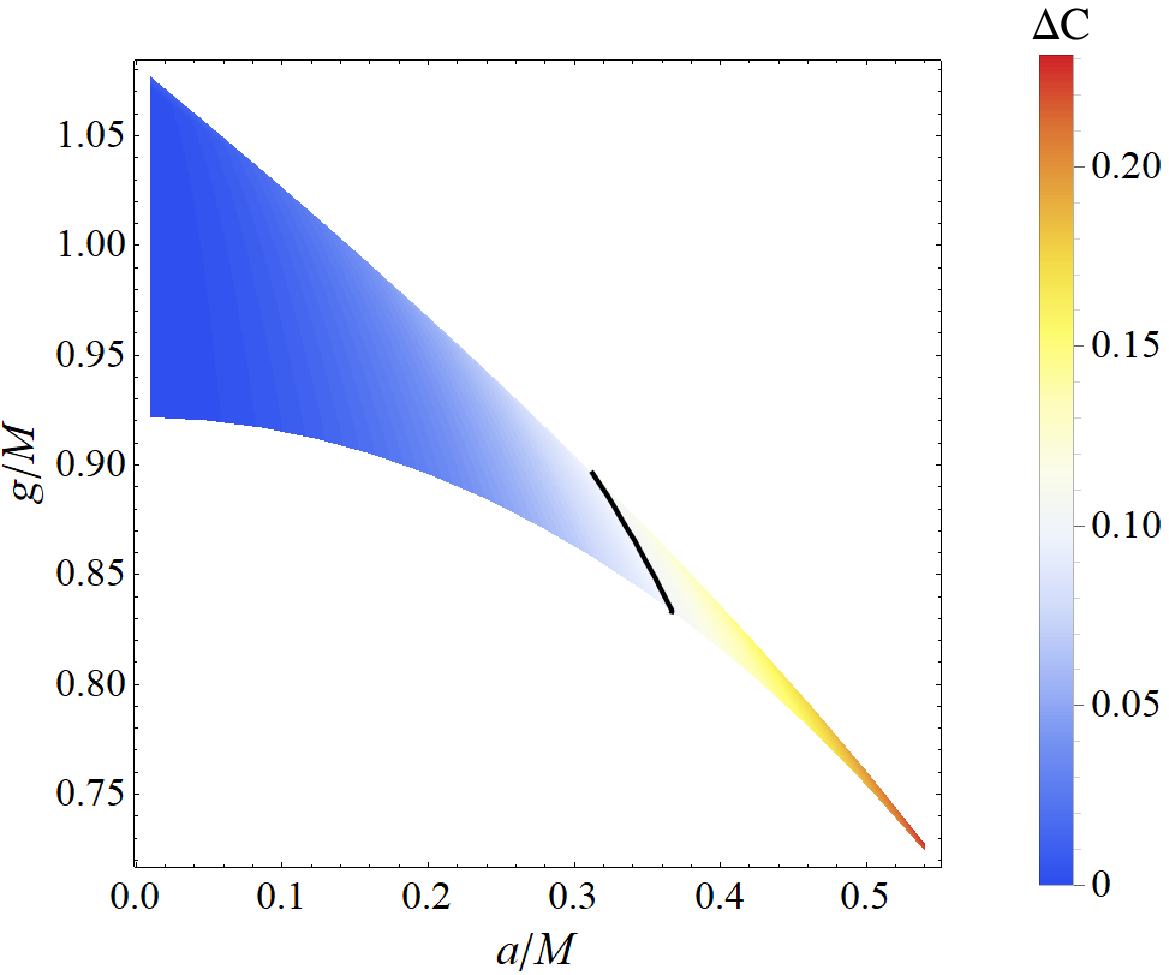}
 \end{tabular}
	\caption{The photon ring circularity deviation observable $\Delta C$ for no-horizon rotating Hayward spacetimes as a function of ($a, g$), $Q=0.0$ (left panel) and  $Q=0.50M$ (right panel). The solid black line is for $\Delta C=0.10$, such that the region on the right-side of the black line is excluded by the measured circularity deviation for the	M87* black hole reported by the EHT, $\Delta C\leq 0.10$.}\label{Hm87}
\end{center}
\end{figure*}
In addition, the rotating Hayward black hole spacetime, as well as all solutions of non-linear electrodynamics theories, are geodesically incomplete; timelike and null geodesics end up at $r=0$ for a finite value of their affine parameter \cite{Lamy:2018zvj}. Unlike the Kerr naked singularity case, we did not extend the spacetime to the $r<0$ due to the de-Sitter core near $r=0$. Thus no light from the negative $r$ region can come to the external observer. 

The circularity deviation observable $\Delta C$ of these shown no-horizon spacetime photon rings ($Q=0.0, g=0.90M, a=0.52M$) and ($Q=0.50M, a=0.43M, g=0.80M$) are, respectively, 0.19242 and 0.13816. Further, in Fig.~\ref{Hm87}, we have shown circularity deviation $\Delta C$ for the closed photon ring of no-horizon rotating Hayward spacetime as a function of ($a,g$), clearly the M87* bound, $\Delta C\leq 0.10$ is satisfied for a finite region of ($a,g$).

\subsection{Nonsingular spacetime}
The nonsingular black hole is a novel class of regular black hole, which in contrast to the Bardeen and Hayward regular black holes, has an asymptotically Minkowski core \cite{Culetu:2014lca,Simpson:2019mud}. 
\begin{figure*}[h!]
	\begin{tabular}{c c}
		\includegraphics[scale=0.8]{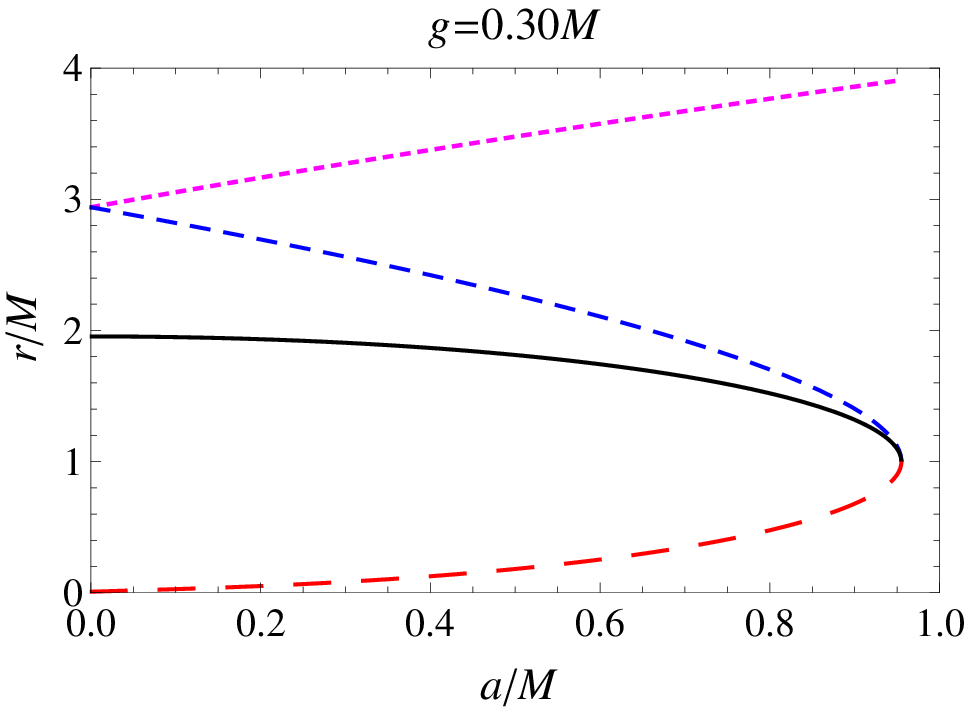}&
		\includegraphics[scale=0.8]{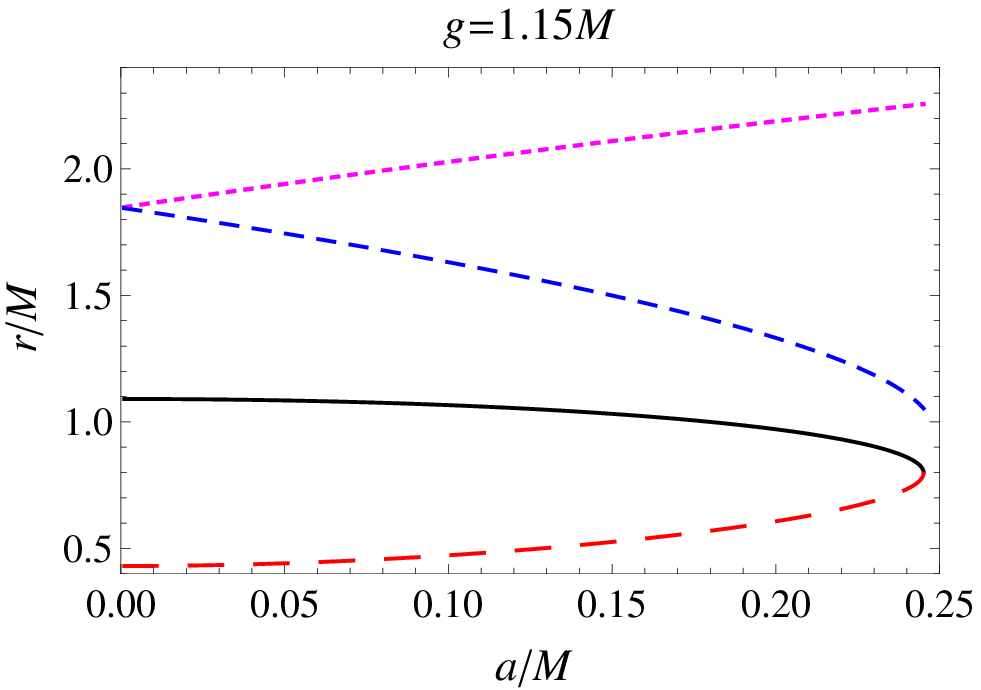}\\
	\end{tabular}
	\caption{Radii of Cauchy horizon $r_-$ (\textit{Red dashed curve}), event horizon $r_+$ (\textit{Black solid curve}),  prograde orbit $r^-_p$ (\textit{Blue small dashed curve}), retrograde orbit $r^+_p$ (\textit{Magenta dotted curve}) for rotating nonsingular black holes.}\label{NSradii}
\end{figure*}
\begin{table*}
	\centering
\begin{tabular}{ |p{1.5cm}|p{1.8cm}|p{1.8cm}|p{1.8cm}|p{1.8cm}|p{2.9cm}| }
	
	\hline
	$g/M$ &  $a_{E}/M$ & $r_{E}/M$ & $r^-_p/M$ &  $r^+_p/M$ & $\delta=(r^-_p-r_{E})/M$  \\
		\hline\hline
		0.0 &  1 &  1 & 1 &  4  &  0 \\
		\hline
		0.10 & 0.995& 0.99998& 0.99998& 3.98999 &  0 \\
		\hline
		0.20 &  0.979999 &0.999803 &0.999803 &3.95985 &  0 \\
		\hline
		0.30 & 0.954984 & 0.999015 & 0.999015 & 3.90921 &  0 \\
		\hline
		0.40 & 0.919914& 0.996948& 0.996948 & 3.83741 &  0 \\
		\hline
		0.50 & 0.874669&  0.992707& 0.992707& 3.74339 &  0 \\
		\hline
		0.60 & 0.818995 &0.98521 &0.98521 &3.62551 & 0 \\
		\hline
		0.70 & 0.752392 & 0.973158 &0.973158& 3.48125&  0 \\
		\hline
		0.80 & 0.673917 & 0.954954& 0.954954 &3.30659 &0 \\
		\hline 
		0.90 & 0.581749& 0.928489& 0.928489&3.09472 &0 \\
		\hline 
		1.00 &0.472045 & 0.89064& 0.89064 &2.83264 &0 \\
		\hline 
		1.10 &0.334606&0.835888&0.884741& 2.48879 &0.0488538 \\
		\hline 
		1.20 &0.108811&0.750887&1.34572&1.89459 &0.548865 \\
		\hline 
		1.21306 &0.0&0.73576&1.61053& 1.61053 &0.874771 \\
		\hline 
	\end{tabular}
	\caption{Horizon radius $r_{E}$, prograde and retrograde photon orbit radii $r^-_p$, $r^+_p$ for various extremal rotating nonsingular black holes. Photon region size decrease with increasing $g$. }\label{NStable}
\end{table*}
\begin{figure*}[h!]
\begin{center}	
	\includegraphics[scale=0.85]{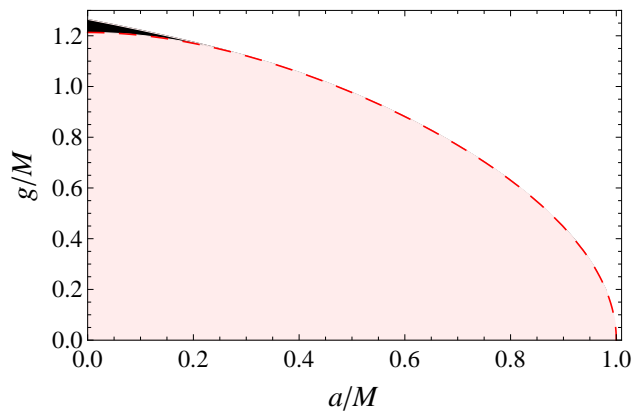}
	\caption{Parameter plane ($a, g$) for the rotating nonsingular spacetime. The dashed red line separates the black hole spacetimes from the no-horizon spacetimes. The no-horizon spacetimes also admit closed photon ring when parameters ($a, g$) lie in the black shaded region.}\label{NS}
\end{center}
\end{figure*}
\begin{figure*}[h!]	
	\begin{tabular}{c c}
		\includegraphics[scale=0.8]{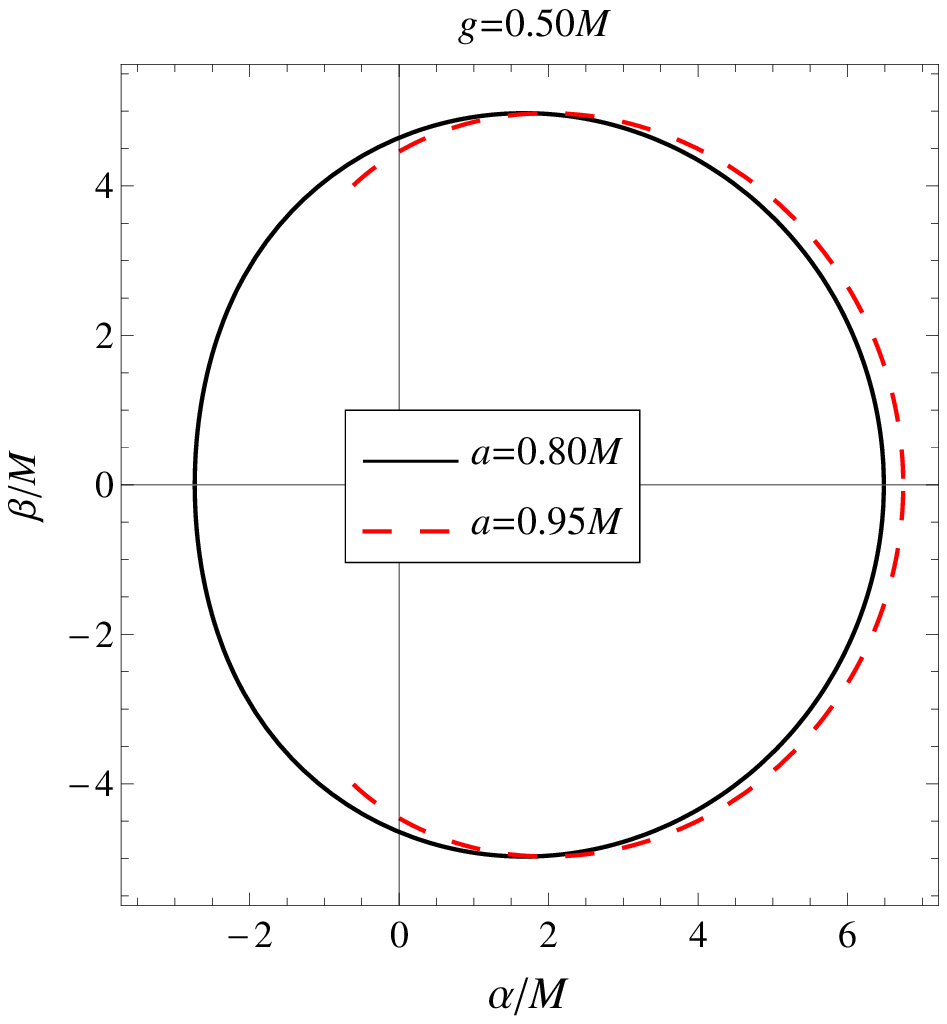}&
		\includegraphics[scale=0.8]{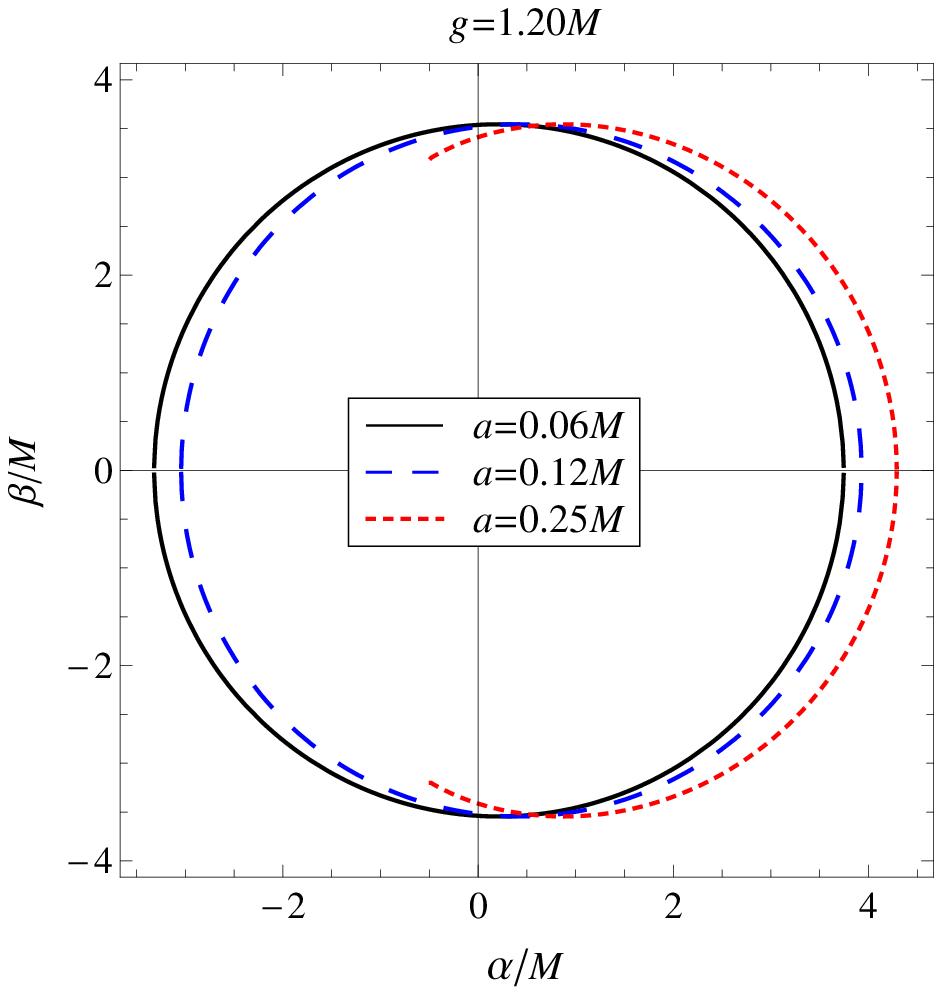}\\
	\end{tabular}
	\caption{Comparison of photon rings for rotating nonsingular spacetimes. Black solid curves correspond to those for black holes, whereas dashed and dotted curves are for the no-horizon spacetimes.}\label{NS1}
\end{figure*}
\begin{figure*}
\begin{center}	
	\includegraphics[scale=0.75]{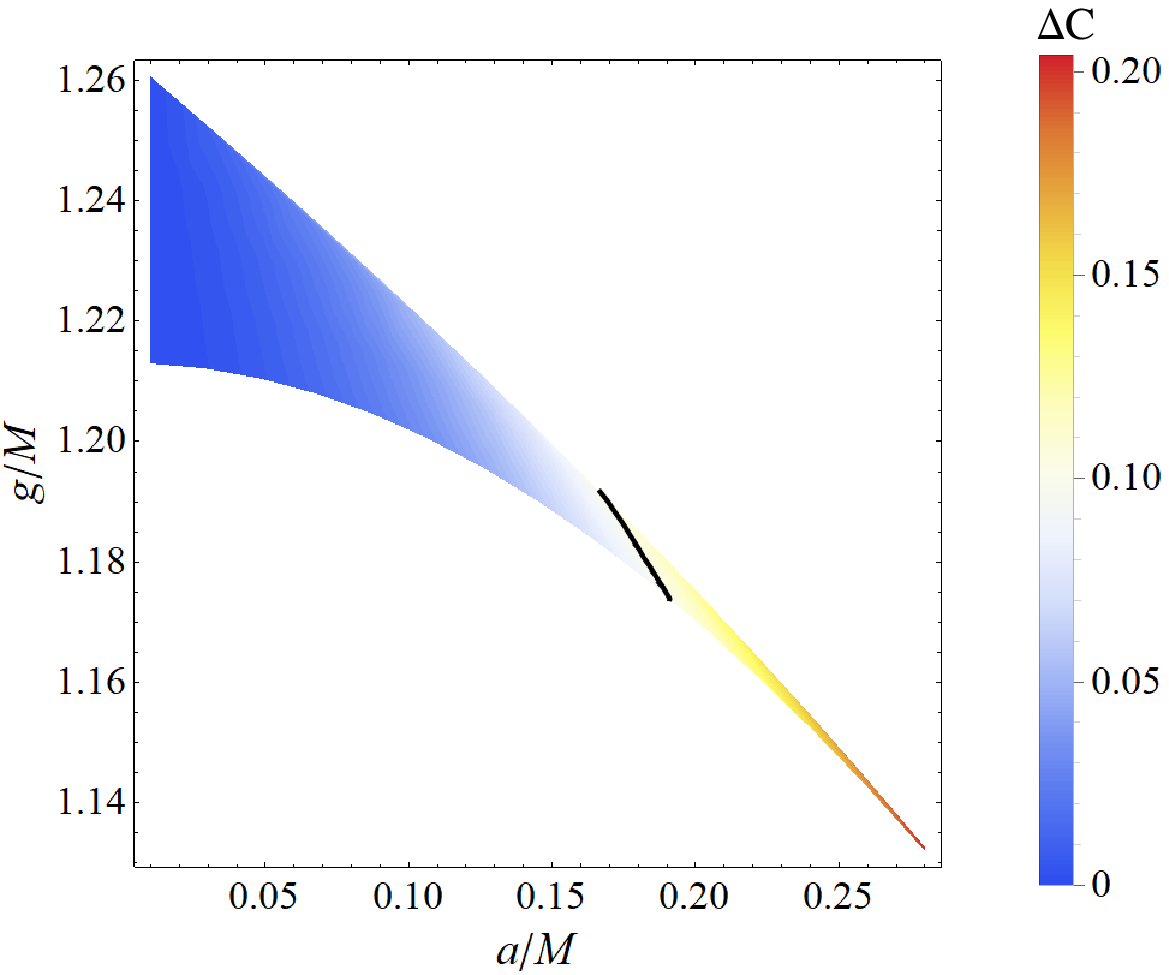}
	\caption{The photon ring circularity deviation observable $\Delta C$ for no-horizon rotating nonsingular spacetimes as a function of ($a, g$). The solid black line is for $\Delta C=0.10$, such that the region on the right-side of the black line is excluded by the measured circularity for the M87* black hole reported by the EHT, $\Delta C\leq 0.10$.}\label{Nm87}
\end{center}
\end{figure*}
These rotating nonsingular or regular black holes also belong to the family of non-Kerr black holes in which the metric tensor resembles with that of the Kerr black hole with mass $M$ replaced by $m(r)=M e^{-g^2/2Mr}$ \cite{Ghosh:2014pba}. 
Further, it may look like a Kerr black hole with a different spin \cite{Kumar:2020yem,Kumar:2018ple}. It turns out that the rotating nonsingular black hole shadows get more distorted and smaller for $g\neq 0$ when compared with the Kerr black hole shadows \cite{Amir:2016cen}. \\
The horizon radii $r_{\pm}$ and photon orbit radii $r_p^{\pm}$ are shown as a function of spin parameter $a$ in Fig.~\ref{NSradii}, and the qualitative behavior is very much similar to those for rotating Bardeen and charged rotating Hayward black holes.
Table \ref{NStable} summarizes the values of $r_{\pm}$ and $r_p^{\pm}$ for various extremal black hole configurations, such as for a given $g$ ($g\leq 1.086M$) and $a=a_E$, $\delta=r_p^--r_E=0$. Unstable circular photon orbits may exist in the no-horizon nonsingular spacetime, viz., for $a=0$, photon prograde orbits exist for $g\leq 1.264523M$ whereas $g_E=1.21306M$. Interestingly, these orbits continue to persist even in the no-horizon rotating spacetimes for suitable values of parameters ($g> 1.086M$) shown as a black shaded region in Fig.~\ref{NS}. This parameter space, admitting the photon orbits, for no-horizon rotating nonsingular spacetime, is comparatively smaller than those for Bardeen and charged Hayward spacetimes. In Fig.~\ref{NS1} the photon ring of nonsingular spacetimes are depicted, and it is evident that the ring for a particular no-horizon spacetime ($a=0.12M, g=1.20M$) has a closed structure with circularity deviation $\Delta C=0.0438986$ and shows resemblance with that for the rotating black hole ($a=0.10M, g=1.20M$). The circularity deviation $\Delta C$ for the no-horizon rotating nonsingular spacetime possessing closed photon ring is shown in Fig.~\ref{Nm87}, clearly the observed bound, $\Delta C\leq 0.10$, for the M87* black hole, is satisfied. Hence, a closed photon ring with a given circularity deviation may correspond to both a rotating black hole or a no-horizon rotating spacetime.

\section{Conclusions}\label{sect5}
In the past decade, there has been a growing interest in the possibility of testing the horizonless compact objects against the Kerr black hole. The bright photon ring, a projection along the photon orbits many times before reaching the observer, has imprints of the spacetime geometry and is anticipated to deduce the exact nature of these objects potentially. In this paper, we have studied the propagation of light rays in three familiar rotating regular no-horizon spacetimes, namely, Bardeen, charged Hayward and nonsingular, and examined and compared their photon rings with those for the Kerr black hole and Kerr naked singularity. We find that rotating regular no-horizon spacetimes (but not always), depending on the values of parameters, possess unstable photon orbits and cast closed photon ring that is very similar to those produced by the Kerr black holes (cf. Figs. \ref{BardeenShadow}, \ref{HayShadow} and \ref{NS1}). Therefore, it is expected that more precise and accurate astrophysical observations can test these no-horizon regular spacetimes against the Kerr hypothesis.

Interestingly, for a given $a$, the existence of photon unstable prograde and retrograde circular orbits in the no-horizon rotating regular spacetimes leads to the closed photon ring only when $g_E<g\leq g_c$, where $g_E$ and $g_c$ for the various spacetimes are given in the Table \ref{table5}.
\begin{table}[h!]
\begin{center}	
	\begin{tabular}{ |c|c|c||c|c|}
		\hline\hline
		\multirow{2}{*}{Spacetimes}  & \multicolumn{2}{c||}{$a=0.10M$} & \multicolumn{2}{c|}{$a=0.20M$} \\	\cline{2-5} 
		& $g_E$  & $g_c$  & $g_E$  & $g_c$ \\
		\hline
		Bardeen&  $0.763332M$      &  $0.816792M$ & $0.744218$ & $0.77257$  \\
			\hline
		\begin{tabular}[c]{@{}l@{}}Charged Hayward\\ ($Q=0$)\end{tabular}
		& $1.05297M$ & $1.164846M$ &   $1.03445M$& $1.108675$ \\ 	\hline
		\begin{tabular}[c]{@{}l@{}}Charged Hayward\\ ($Q=0.50M$)\end{tabular} 
		& $0.915311M$& $1.026494M$ & $0.895853M$ & $0.967483M$  \\	\hline
		Nonsingular&  $1.2020M$    & $1.222461M$ & $1.17029M$ & $1.17533704M$  \\
		\hline\hline     
	\end{tabular}
\caption{Values of $g_E$ and $g_c$, for a given $a$ for existence of closed photon ring.}
\end{center}\label{table5}
\end{table}
The bounds on the spin parameter, for which Bardeen, Hayward ($Q=0.0$ and $Q=0.50M$) and nonsingular no-horizon spacetime possess closed photon ring ($g_E<g\leq g_c$) are, respectively, $0\leq a\leq 0.5730M$, $0\leq a\leq 0.7216M$, $0\leq a\leq 0.6688M$ and $0\leq a\leq 0.3563M$. When the values of $a$ and $g$ exceed the obtained bounds, the prograde orbits disappear and the ring reduces into an open arc (cf. Figs. \ref{BardeenShadow}, \ref{HayShadow} and \ref{NS1}). This result is in contrary to the Kerr no-horizon spacetimes (naked singularities) which never form closed photon ring, but is in accordance with the Kerr-Newman no-horizon (naked singularities) which have similar features. 

Our study suggests that the structure of photon ring for no-horizon rotating spacetime may resemble those for the Kerr black hole, therefore it is of theoretical and astrophysical relevance in light of ongoing EHT observations. The key indication that follows from our analysis is that the mere existence of a photon ring does not by itself imply the presence of a black hole or event horizon.

In conclusion, we have examined photon rings of rotating regular no-horizon spacetimes, compared them with those for the black holes, and saw whether the two are clearly distinguishable. Interestingly, while black holes always cast a closed photon ring, rotating regular no-horizon spacetimes may or may not. Therefore, while black holes imply the existence of the photon ring enclosing the dark shadows, the converse is not valid. A closed photon ring can also be produced by rotating regular no-horizon spacetimes as well. Further, while Kerr naked singularity spacetimes never cast a closed ring, Kerr-Newman and three rotating regular no-horizon spacetimes, for certain parameter values,  do generate a closed ring. Therefore, besides black holes and Kerr-Newman naked singularities, rotating regular no-horizon spacetimes are also capable of forming closed photon ring.

Further, the ring circularity deviation observable $\Delta C$ for the rotating no-horizon spacetimes, for certain values of parameters ($a, g$), may satisfy the observational bound deduced for the M87* black hole, i.e., $\Delta C\leq 0.10$. Other interesting phenomena related to the no-horizon rotating spacetimes and the shadows in the presence of accretion disk are subjects of our future investigation.

\section{Acknowledgment}
S.G.G. would like to thank  SERB-DST for the ASEAN project IMRC/AISTDF/CRD/2018/000042  and also to IUCAA, Pune for the hospitality while this work was being done. R. K. thanks UGC, Government of India for financial support through SRF scheme.
\appendix
\section{Kerr-Newman naked singularity}
The Kerr-Newman black hole is described by metric (\ref{rotmetric}) with mass function \cite{Newman:1965my}
\begin{equation}
m(r)=M-\frac{Q^2}{2r}.
\end{equation}
\begin{figure*}[b!]
	\begin{tabular}{c c}
		\includegraphics[scale=0.75]{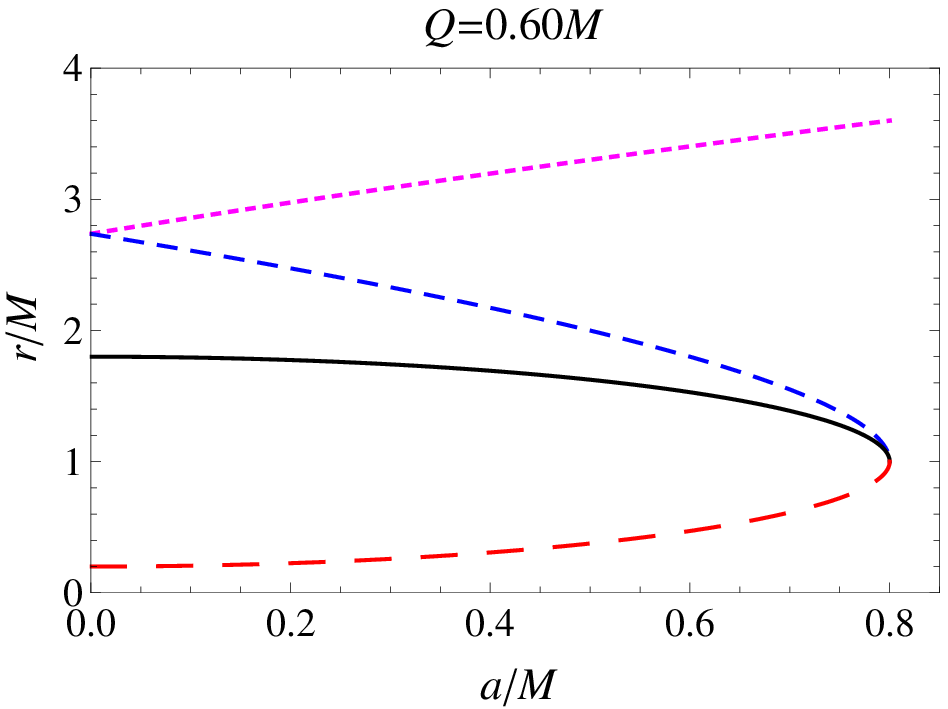}&
		\includegraphics[scale=0.75]{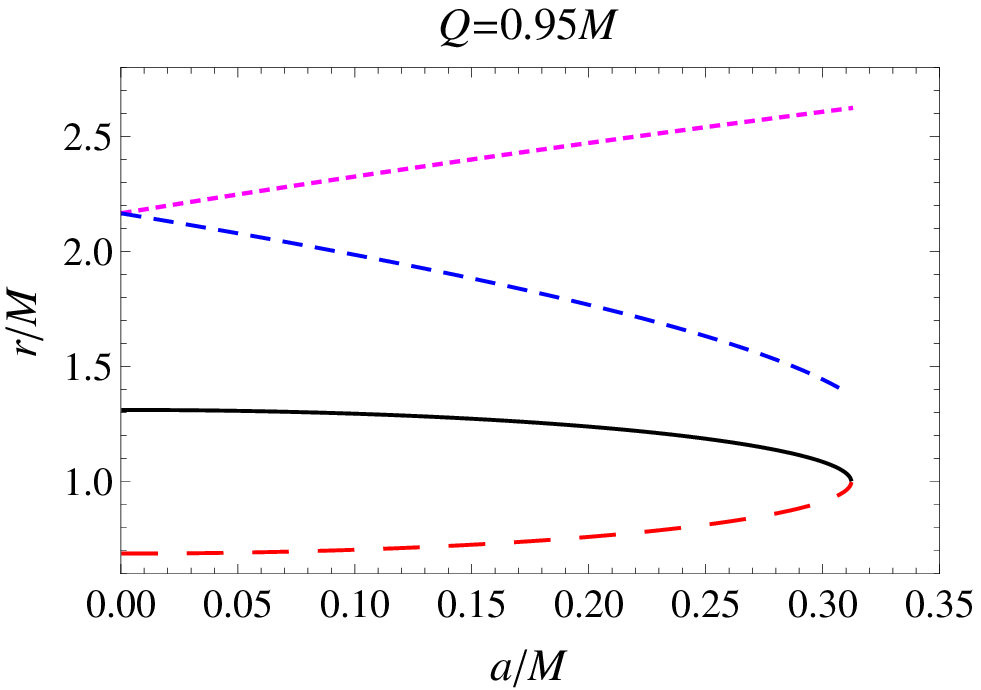}\\
	\end{tabular}
	\caption{Radii of Cauchy horizon $r_-$ (\textit{Red dashed curve}), event horizon $r_+$ (\textit{Black solid curve}),  prograde orbit $r^-_p$ (\textit{Blue small dashed curve}), retrograde orbit $r^+_p$ (\textit{Magenta dotted curve}) for Kerr-Newman black holes.  }\label{KNradius}
\end{figure*}
\begin{table}
	\centering
	\begin{tabular}{ |p{1.5cm}|p{1.8cm}|p{1.8cm}|p{1.8cm}|p{1.8cm}|p{2.9cm}| }
		
		\hline
		$Q/M$ &  $a_{E}/M$ & $r_{E}/M$ & $r^-_p/M$ &  $r^+_p/M$ & $\delta=(r^-_p-r_{E})/M$  \\
		\hline\hline
		0.0 &  1 &  1 & 1 &  4  &  0 \\
		\hline
		0.10 &  0.994987 &  1 & 1 & 3.98997 &  0 \\
		\hline
		0.20 &  0.979795 & 1 & 1 & 3.95959 &  0 \\
		\hline
		0.30 &  0.953939 &  1 & 1 & 3.90788 &  0 \\
		\hline
		0.40 &  0.916515 &  1 & 1 & 3.833033 &  0 \\
		\hline
		0.50 &  0.866025 &  1 & 1 & 3.73205 &  0 \\
		\hline
		0.60 &  0.80 & 1 & 1 & 3.60 & 0 \\
		\hline
		0.70 &  0.714143 &  1 & 1  & 3.42829 &  0 \\
		\hline
		0.80 &  0.60 &  1 & 1 & 3.20 & 0\\
		\hline 
		0.90 &  0.435890 &  1 & 1.12822 & 2.87178 & 0.12822 \\
		\hline 
		0.95 &  0.312249 &  1 & 1.37550 & 2.62450 & 0.37550 \\
		\hline 
		0.99 &  0.0141067 &  1 & 1.71787 & 2.28213 & 0.71787 \\
		\hline 
	\end{tabular}
	\caption{Table summarizing the values of extremal horizon radius $r_{E}$, prograde and retrograde photon orbit radii, respectively, $r^-_p$ and $r^+_p$ of Kerr-Newman black holes. With increasing $Q$ the size of the photon region decreases.}\label{KNtable}
\end{table}
\begin{figure*}
\begin{center}
	\includegraphics[scale=0.85]{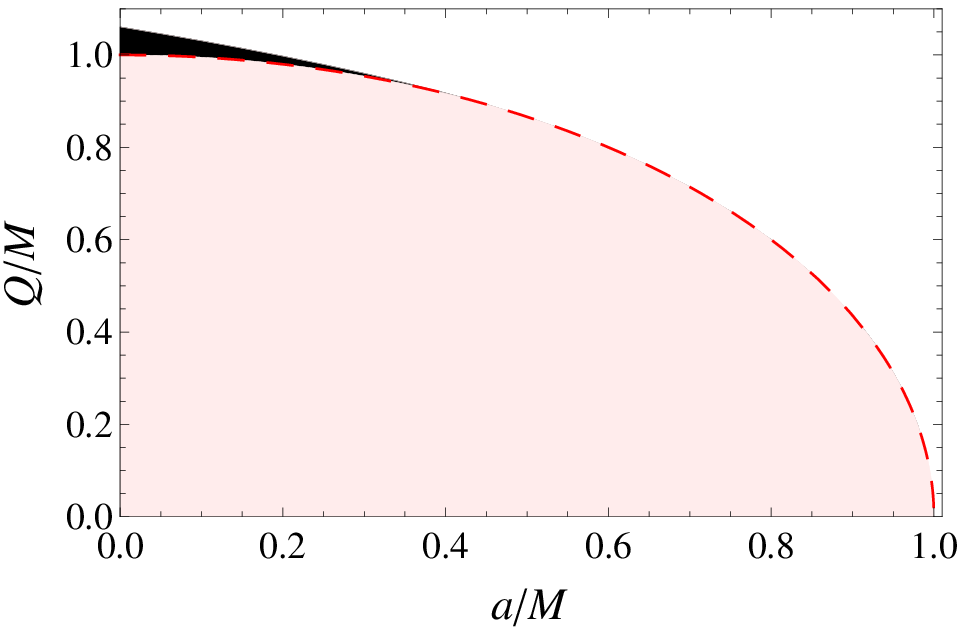}
	\caption{Parameter plane ($a, Q$) for the Kerr-Newman spacetime. The red dashed line separates the black hole spacetimes from the naked singularity spacetimes. The naked singularity spacetimes also admit closed photon ring when parameters ($a, Q$) lie in the black shaded region. }\label{KN}
\end{center}	
	\begin{tabular}{c c}
		\includegraphics[scale=0.8]{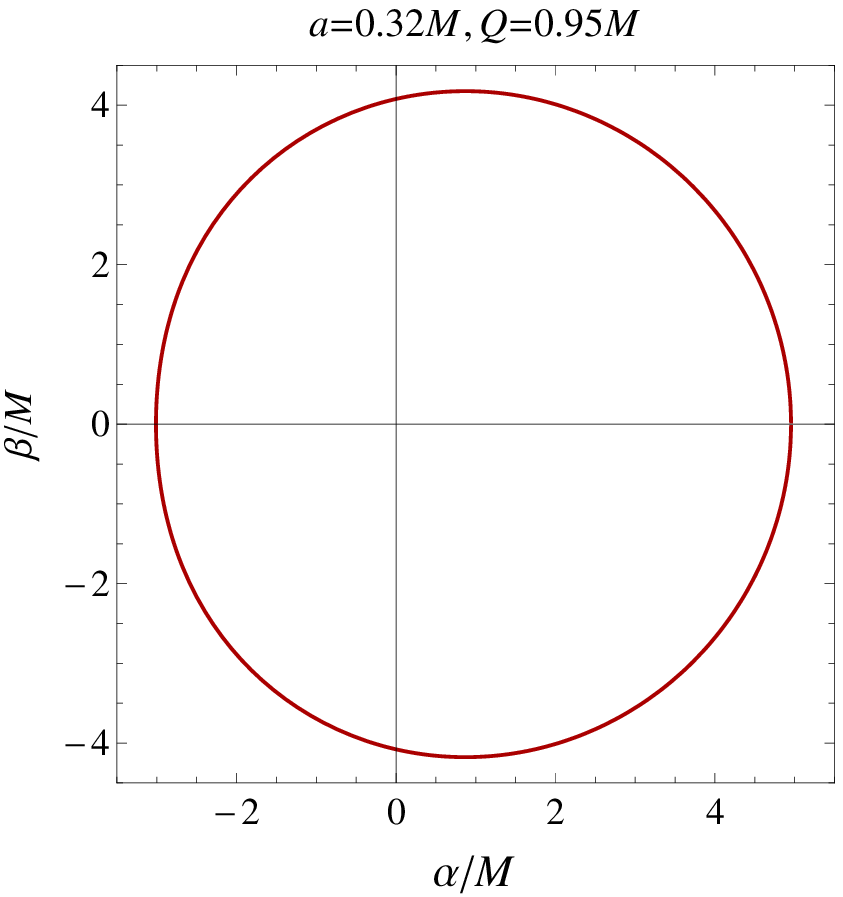}&
		\includegraphics[scale=0.8]{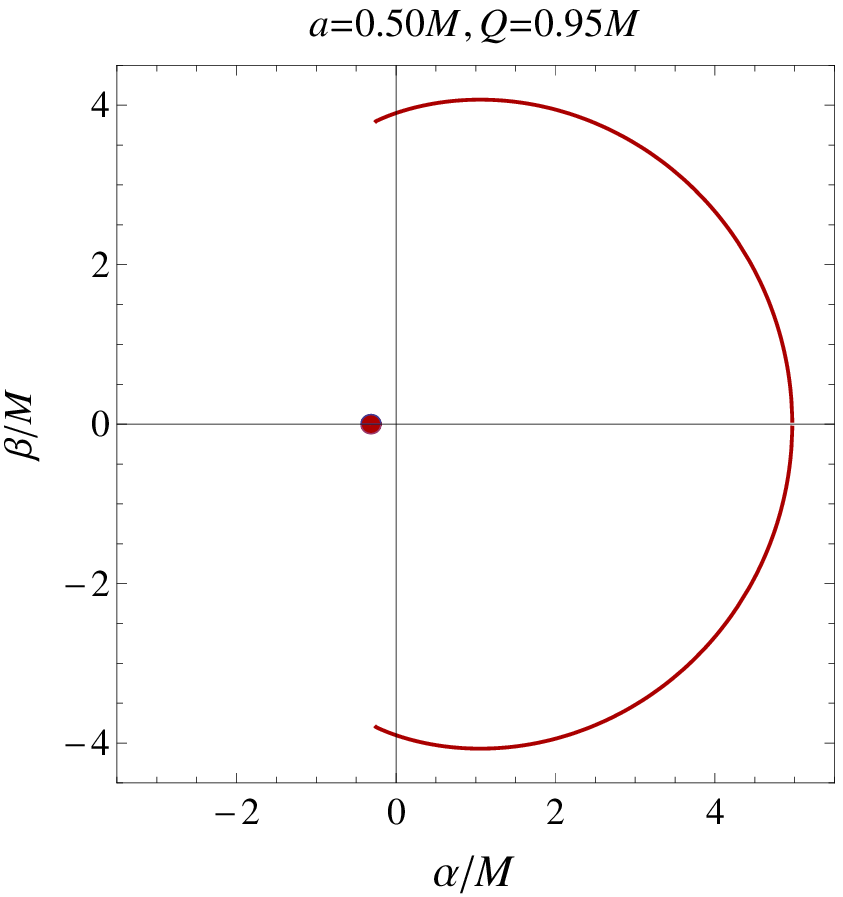}\\
	\end{tabular}
	\caption{Photon ring around the Kerr-Newman naked singularities for different values of $a$ and inclination angle $\theta_o=\pi/3$.}
	\label{KNmain}
\begin{center}	
	\includegraphics[scale=0.73]{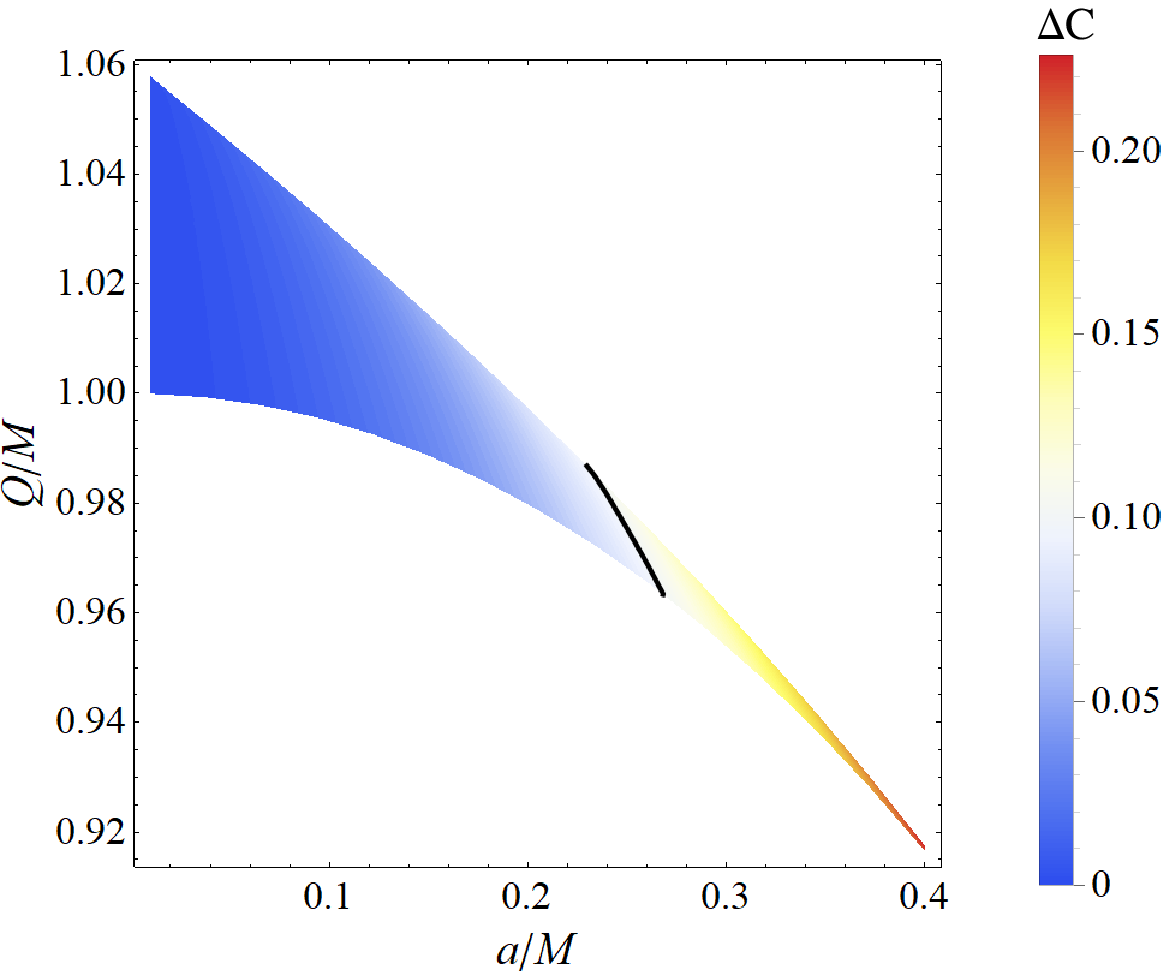}
	\caption{The photon ring circularity deviation observable $\Delta C$ for Kerr-Newman naked singularities as a function of ($a, g$).  The black solid line is for $\Delta C=0.10$, such that region on the right-side of the black line is excluded by the measured circularity for the M87* black hole reported by the EHT, $\Delta C\leq 0.10$.}\label{KNm87}
\end{center}
\end{figure*}
The behavior of horizon radii $r_{\pm}$ and photon orbit radii $r_p^{\pm}$ with varying spin parameter $a$ and for different values of $Q$ is depicted in Fig.~\ref{KNradius}, which shows that prograde orbits radii $r_p^-$ decrease and retrograde orbits radii $r_p^+ $ increase with $a$. Table~\ref{KNtable} summarize these values for various extremal black hole configurations, such that for given $Q$ ($Q\leq 0.8661M$) there exists $a=a_E$, for which prograde orbits coincide with the extremal horizon radii, i.e., $\delta=r_p^--r_E=0$. Whereas for $Q>0.8661M$ and $a=a_E$, $\delta=r_p^--r_E\neq 0$ and $\delta$ further increases with $Q$. Thus, for sufficiently large values of electric charge $Q\, (>0.8661M)$ and small rotation parameter, prograde photon orbits can persist for naked singularity spacetimes.
Kerr-Newman black hole spacetimes posses retrograde and prograde unstable photon orbits, moreover, these orbits also exist in the naked singularity spacetime with finite parameter space ($a,Q$) shown as a black shaded region in Fig.~\ref{KN}, viz. for $a=0.10M$, $0.99498M<Q\leq 1.0305201M$. Outside the parameter space, prograde orbits disappear. This leads to an important consequence that the loci of photon orbits in the Kerr-Newman naked singularity spacetimes may form a closed photon ring, which is strikingly different from that for the Kerr naked singularity (cf. Figs.~\ref{KerrNS}, \ref{KNmain}). The central small dark spot in Fig.~\ref{KNmain} corresponds to the admissible photon orbits in the negative $r$ region. It is clear that only slowly rotating no-horizon spacetimes ($0\leq a\leq 0.4998M$) can cast closed photon ring (cf.  Table \ref{KNtable} and Fig.~\ref{KN}). In Fig.~\ref{KNm87}, ring circularity deviation observable  $\Delta C$ is plotted as a function of ($a, Q$) for the Kerr-Newman naked singularity spacetimes causing the closed photon ring. It is evident that these photon rings for the naked singularities can satisfy the observational bound, $\Delta C=0.10$, for the M87* black hole.

\noindent
\end{document}